**Optimizing the Accuracy of Viscoelastic Characterization with AFM Force-Distance Experiments in the Time and Frequency Domains**

*Marshall R. McCraw, Berkin Uluutku, Halen D. Solomon, Megan S. Anderson, Kausik Sarkar, and Santiago D. Solares\**

*Department of Mechanical and Aerospace Engineering, The George Washington University School of Engineering and Applied Science, Washington, District of Columbia, USA*

\*Corresponding Author: Santiago D. Solares – ssolares@gwu.edu

**Abstract**

We demonstrate that the method of characterizing viscoelastic materials with Atomic Force Microscopy (AFM) by fitting analytical models to force-distance (FD) curves often yields conflicting and physically unrealistic results. Because this method involves specifying a constitutive time-dependent viscoelastic model and then fitting said model to the experimental data, we show that the inconsistencies in this method are due to a lack of sensitivity of the model with respect to its parameters. Using approaches from information theory, this lack of sensitivity can be interpreted as a narrowed distribution of information which is obtained from the experiment. Furthermore, the equivalent representation of the problem in the frequency domain, achieved via modified Fourier transformation, offers an enhanced sensitivity through a widening of the information distribution. Using these distributions, we then define restrictions for the timescales which can be reliably accessed in both the time and frequency domains, which leads to the conclusion that the analysis of experiments in the time domain can frequently lead to inaccuracies. Finally, we provide an example where we use these restrictions as a guide to optimally design an experiment to characterize a polydimethylsiloxane (PDMS) polymer sample.

1. **Introduction**

The field of mechanobiology has provided a number of powerful insights into the rich behaviour of soft, biological materials[1–6]. Within this field, colloid based rheology measurements have allowed researchers to probe both the passive and active time-dependent responses of biological materials across a large range of timescales, making it possible to directly measure the contribution of individual bio-polymers to a cell's global behaviour[7–15]. Although these micro-rheological techniques are among the most prevalent in the mechanobiology community, Atomic

Force Microscopy (AFM) and, more generally, Scanning Probe Microscopy (SPM), have become popular alternatives due to their uniquely versatile range of *in vitro* measurements and non-destructive nature[2,3,16–21]. Although these methods are typically incapable of achieving a similarly broad characterization of the viscoelastic timescales of a material, attempts have been made to close this gap[22–24]. Still, the most common viscoelastic characterization technique among AFM practitioners is the force-indentation (or force-distance, FD) experiment. Due to its simplicity and short duration, FD experiments are an attractive option for those seeking high throughput data acquisition to probe the heterogeneities inherent in biological materials[19,25]. Despite previous attempts to leverage the wide-band nature of thermal excitations in AFM based measurements, the traditional method used to characterize the viscoelastic properties of a material is still the force-indentation (or force-distance FD) experiment, due to its simplicity and short duration – allowing high throughput data acquisition.

Within AFM, other methods such as band excitation, stress relaxation (creep testing), and dynamic mechanical analysis can also be performed; however, such experiments typically take longer, thus limiting the amount of data that can be obtained from an experiment when measuring sensitive, active materials like cells, for example[19,26,27].

In force-indentation experiments, the surface of the material is indented with the AFM probe while the indentation depth $h$ and the resulting interaction force $f$ are recorded in a way that is analogous to macro-scale tensile and compression test experiments. Using contact mechanics, one can relate the force and indentation in terms of a constitutive stress-strain equation which, for the case of a linear viscoelastic material, is generally a convolution integral as shown in Eqn. 1 and 1a[28–35]. Here, the terms $\alpha$ and $\beta$ depend on the geometry of the AFM probe and can be found in further sources[19,28,30–32].

$$\sigma(t) = \int du \ Q(t-u)\varepsilon(u) \tag{1}$$

$$\frac{f(t)}{\alpha} = \int du \ Q(t-u)h^\beta(u) \tag{1a}$$

$$Q(\omega) = \frac{\sigma(\omega)}{\varepsilon(\omega)} = \frac{f(\omega)}{\alpha \ h^\beta(\omega)} \tag{2}$$

In the state-of-the-art, the viscoelastic modulus $Q$ can then be obtained by first assuming a functional form for $Q$ and then fitting the specific constitutive equation to the obtained force-indentation or stress-strain data in what we will refer to as the time domain approach[18,19,36–40]. Alternatively, one can use certain integral transforms, for which the convolution theorem is valid, to 'undo' the convolution and directly obtain the modulus from the transformed force-indentation data without prescribing a functional form to $Q$ as seen in Eqn. 2 (here, we will use the modified Fourier transform with the notation $f(\omega)$ denoting the transform of $f$)[41,42]. While this 'frequency domain approach' allows an independence from model prescription, obtaining a model-based parameterization is often necessary for communication and comparison purposes. Thus, just as is done in the time domain approach, models for $Q(\omega)$ can be chosen and then fit to the transformed data.

While these methods offer straightforward solutions for obtaining parameterized descriptions of viscoelastic materials, their accuracy and reliability has recently been questioned[40,43]. Specifically, the time domain approach often yields unphysical or even conflicting values of the model parameters which depend on the initialization of the fitting algorithm[44]. Furthermore, a recent work by Vemaganti *et al* demonstrates that relaxation times can only be reliably obtained from stress relaxation experiments if they are less than 20% of the total experiment length[45]. Despite these potentially major inconsistencies, the results obtained from the time domain approach often are in close agreement with the data and require a close inspection to validate their physical sensibility. However, such an approach would be completely impractical in high throughput applications which often involve 1000's of separate force curves within a single measurement.

## 2. Theoretical Background

Although not an exhaustive list, researchers tend to choose from Maxwell, Kelvin-Voigt, power-law, and fractional calculus-based models for the parametrization of the viscoelastic modulus[18,19,36–38,46]. Here, we will focus specifically on the generalized Maxwell model $Q_{GM}$ seen in Fig. 1b and defined by Eqn. 3 and 3a for the time and frequency domains, respectively (note that the modified Fourier domain representation is used in Eqn. 3a; further details can be found in the appendix). As seen in Fig. 1b, the generalized Maxwell model is comprised of a series of $N$ 'Maxwell arms' in parallel with a single elastic element $G_e$. Each arm has a spring element contributing an elasticity $G_n$ and a dashpot (damper) element contributing a viscosity $\mu_n$. The ratio of these arm components gives a characteristic relaxation time $\tau_n = \mu_n/G_n$ which governs the rate at which the stress in the $n^{th}$ arm relaxes, with larger values corresponding to slower stress relaxation and vice-versa. By increasing the number of Maxwell arms $N$, one can describe a material with an intricate relaxation process. For example, a two-armed generalized Maxwell model was used to describe the 'fast' and 'slow' cytoskeletal rearrangement of cancer cells from stress-relaxation experiments[47].

$$Q_{GM}(t) = G_e \delta(t) + \sum_{n}^{N} \left[ G_n \delta(t) - \frac{G_n}{\tau_n} e^{-\frac{t}{\tau_n}} \right] \tag{3}$$

$$Q_{GM}(\omega) = G_e + \sum_{n}^{N} \left[ G_n - \frac{G_n}{1 + \frac{\tau_n}{\Delta t}\left(1 - \frac{e^{-i\omega}}{1.001}\right)} \right] \tag{3a}$$

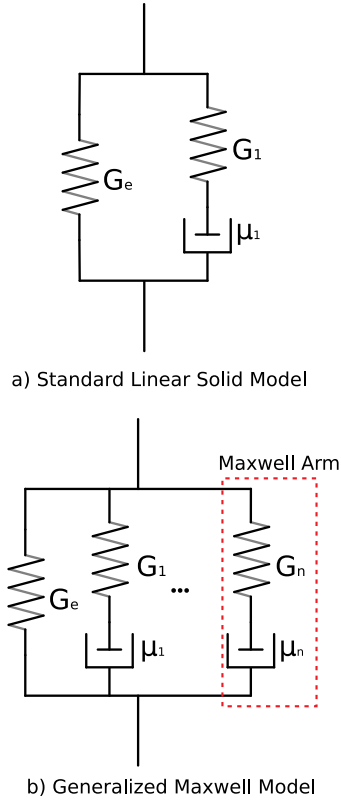

**Figure 1:** a) illustrations of a Standard Linear Solid (SLS) model and b) a generalized Maxwell (GM) model comprised of an arbitrary number of Maxwell arms, seen in dashed red.

As previously mentioned, we have found that the behavior of the generalized Maxwell model is relatively insensitive to changes in its parameter set $\boldsymbol{\theta}$ (i.e., $G_e, G_1, \tau_1$). In the time domain, for instance, while a model fitted to a force-indentation curve may agree with the true data obtained from an experiment, the fitted values of $\boldsymbol{\theta}$ may vary significantly compared to the true values for the material. Such discrepancies are demonstrated in Fig. 2a where 100 generalized Maxwell models were fit to a simulated force-indentation curve (seen in blue). Although there are a few fits (seen in grey) which visually disagree with the true force-indentation data, the average (seen in red) follows the data quite well. If this were representative of a typical characterization experiment of the simulated material, the averaged data would be considered to have successfully described the material due to its agreement with the data. However, the discrepancies in the resulting fitted $\boldsymbol{\theta}$'s become apparent when plotting the storage and loss moduli (real and imaginary parts of $Q(\omega)$) for the fits, their average, and the data as seen in Fig. 2b and c, respectively. Here, we see that values of the relaxation times (inverse of the frequency index of the peak in the loss modulus) disagree by 1-2 orders of magnitude between the fits and the data. Furthermore, as models with $N$ between 1 and 4 were fit to the data, models with different numbers of relaxation times seem to equally describe the same force-indentation behavior in the time domain, despite the qualitative differences in the physics. As a result of this insensitivity to the number of relaxation times, the common practice for materials to be fitted with an arbitrary $N$, raises questions regarding overfitting the data and regarding fitting the data to incorrect physical behaviors. Although the comparison of the storage and loss moduli of the fits and the material seems to allow a more accurate assessment of the fit accuracy, such data is not directly obtained from the time domain fitting approach, thus making this assessment impossible if one fits in the time domain. However, the results suggest that the frequency domain approach should offer an advantage in more accurately parameterizing materials from force-indentation data.

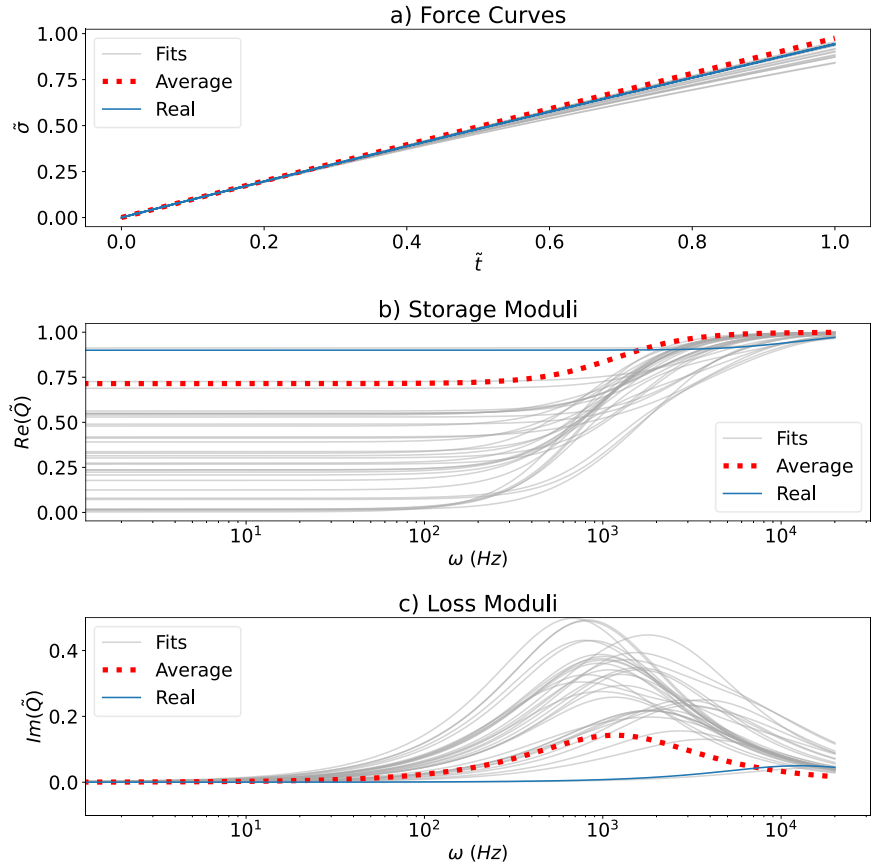

**Figure 2**: a) 100 randomly initialized (N between 1 and 4) fit attempts (grey) to a simulated force-indentation curve (N=1) (blue) with the average behavior of the fit attempts shown in dashed red. b) Storage moduli for each parameter set obtained from fitting and averaging. c) Loss moduli for each parameter set obtained from fitting and averaging.

3. Analysis
    a. **Insensitivity of the $l^2$ Norm**

To compare the capabilities of these two approaches in determining an accurate $\boldsymbol{\theta}$ for a material, we will first use simplified, theoretical cases, treating force-indentation curves as linear strain inputs $h^\beta(t) = \varepsilon(t) \approx \varepsilon_0\, t$. While the indentation in AFM experiments is not typically linear, the 'strain' factor $h^\beta$ in the contact mechanics models often closely follows a line as seen in the technical appendix. With this simplification, the stress-strain behavior for the linearized force-indentation model can be obtained by solving the Volterra integral in Eqn. 1 for the given strain $\varepsilon(t) = \varepsilon_0\, t$ and modulus $Q(t) = Q_{GM}(t)$. We further express both the time $t$ and the relaxation times $\tau_n$ as fractions of the full experiment length $T$ as well as normalize each arm elasticity $G_n$ by the glassy (or instantaneous) modulus $G_g = G_e + \sum G_n$ to remove the dependence on $G_e$ (the equilibrium modulus) and thus, reduce the dimensionality of $\boldsymbol{\theta}$. A similar dimensional analysis was performed for the modified Fourier domain representation of the modulus, yielding $\widetilde{Q}(\omega)$ where $\Delta t$ denotes the sampling timestep of the experiment and 1.001 is the typical radial term used in the discrete modified Fourier transformation of FD curves[42]. One should note that an equivalent non-dimensionalization could be performed by scaling the timescales by the inverse of the strain rate, $\varepsilon_0$. Thus, the findings that are obtained from the analysis of the equations scaled by $T$ are equivalent to those scaled by $1/\varepsilon_0$. The resulting dimensionless equations for the corresponding observables in both the time (stress $\tilde{\sigma}$) and frequency (modulus $\widetilde{Q}$) domains are given in Eqn. 4 and 4a; however, refer to the technical appendix for more thorough derivations. Using these models, simulated datasets $\tilde{\sigma}_{obs}$ and $\widetilde{Q}_{obs}$ can be generated for various values of $\boldsymbol{\theta}$ with the addition of Gaussian white noise with a standard deviation $s$ equal to 0.1% of the full scale of the data. Hence, the simulated datasets are represented as $\tilde{\sigma}_{obs} = \tilde{\sigma} + P_N(0, s^2)$ and $\widetilde{Q}_{obs} = \widetilde{Q} + P_N(0, s^2)$, where $P_N(0, s^2)$ denotes a normal distribution sample with average equal to zero and variance $s^2$. Then, the $\boldsymbol{\theta}$ which describes the simulated data can be obtained through fitting the relevant model by minimization of the $l^2$ norm between the data ($\tilde{\sigma}_{obs}$ or $\widetilde{Q}_{obs}$) and the model 'prediction' ($\tilde{\sigma}$ or $\widetilde{Q}$), as defined in Eqn. 5 and 5a.

$$\tilde{\sigma}(t) = \frac{f(t)}{\alpha \, \varepsilon_0 \, G_g \, T} = \tilde{t} + \sum_n^N \widetilde{G}_n \tilde{\tau}_n \left(1 - e^{-\frac{\tilde{t}}{\tilde{\tau}_n}}\right) - \widetilde{G}_n \tilde{t} \tag{4}$$

$$\widetilde{Q}(\omega) = \frac{Q(\omega)}{G_g} = 1 - \sum_n^N \frac{\widetilde{G}_n}{1 + \frac{\tilde{\tau}_n}{\Delta t}\left(1 - \frac{e^{-i\omega}}{1.001}\right)} \tag{4a}$$

$$l_t^2(\boldsymbol{\theta}) = \sum_{\tilde{t}} [\tilde{\sigma}(\tilde{t}, \boldsymbol{\theta}) - \tilde{\sigma}_{obs}(\tilde{t})]^2 \tag{5}$$

$$l_\omega^2(\boldsymbol{\theta}) = \sum_{\omega} \left[\widetilde{Q}(\omega, \boldsymbol{\theta}) - \widetilde{Q}_{obs}(\omega)\right]^2 \tag{5a}$$

The successful identification of $\boldsymbol{\theta}$ from a dataset is dependent on the structure of the $l^2$ norm; for example, if the norm has multiple local minima, an optimization algorithm may falsely identify one of these points as the true minimum. Furthermore, if the surface of the $l^2$ norm encompasses regions of $\boldsymbol{\theta}$ for which the values of the norm are below the machine precision level, the computer will not be able to reliably distinguish between $\boldsymbol{\theta}$'s in these areas and the algorithm will also be unable to continue advancing towards the minimum[48].

In a mathematically simpler power-law viscoelastic model, it was recently demonstrated that the structure of the $l^2$ norm contained many local minima which resulted in frequent misidentifications of $\boldsymbol{\theta}$[44]. In Fig. 3a and 3b, we show the surfaces of the norm for both the time $l_t^2$ and frequency $l_\omega^2$ domain models for a wide range of parameter space around the true minimum ($\widetilde{G}_{1_{obs}} = 0.5, \tilde{\tau}_{1_{obs}} = 0.5$). Although these two-dimensional models do not have any local minima, the regions of flattened topography with values below the machine precision level introduce the possibility of stagnant optimization algorithms. Specifically, a large valley formation occurs around the minimum of the time domain model as seen in Fig. 3a. This valley encompasses a large range of possible values for the elasticity $\widetilde{G}_1$ and relaxation time $\tilde{\tau}_1$, especially considering the normalization of these parameters, thus implying that parameters in this range offer no discernible change in the behavior of the model in the context of a force-indentation curve, from the perspective of the computer[48]. The variability of the fit results seen in Fig. 2 and the sensitivity to the initialization of the optimizer can then be understood as outcomes of this flatness in the $l^2$ surfaces.

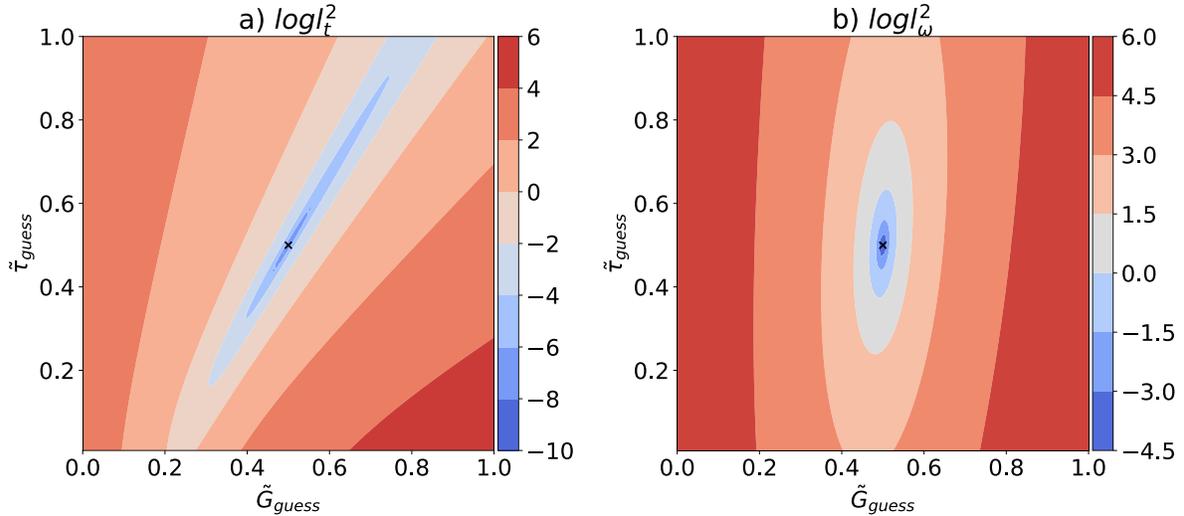

**Figure 3**: Surfaces of the $l^2$ norm for both the a) time and b) frequency domain models for a material with $\tilde{G}_{1_{obs}} = 0.5$ and $\tilde{\tau}_{1_{obs}} = 0.5$ (true values indicated on the plots with an X).

As the valley in the $l_t^2$ norm extends towards larger values of $\tilde{\tau}_1$, one can expect that arms with relaxation times that are comparable to the length of the experiment or close to the inverse strain rate ($\tilde{\tau}_n$ close to 1) will contribute a negligible amount to the model behaviour (since they have negligible effect on the norm). The functional form of the model in the time domain permits such a behavior as the term $1 - e^{-\tilde{t}/\tilde{\tau}_n}$ will remain small for most of the experiment. As the arm behavior is further scaled by the product of the modulus and the relaxation time $\tilde{G}_n\tilde{\tau}_n$, then, in the case where the relaxation time is large, the behavior of the arm is dominated by this product, allowing for errors in $\tilde{\tau}_n$ to be offset by inversely proportional errors in $\tilde{G}_n$, thus explaining the valley formed in the $l_t^2$ norm.

The norm in the frequency domain offers a contrast to the time domain norm, as seen in Fig. 3b. Here, a single, well-defined minimum in the parameter space is surrounded by a convex, highly curved surface, thus indicating that the model in the frequency domain is more amenable to successful optimization attempts, even in the presence of considerable noise. Furthermore, the region of the parameter space that is below the machine precision level is confined to the area immediately around the true minimum, therefore, the problem of optimizers being unable to advance before reaching the true minimum is not present.

Although it may seem counterintuitive that these two representations of the same physics have differing behaviors when optimizing a fit, the finding is of a similar essence as the use of integral transforms in solving ordinary differential equations (ODEs). In the case of ODEs, certain integral transforms change the functional form of the equation allowing direct algebraic solutions to the equations. Here, the integral transforms change the functional form of the model in the time domain to a function that has greater sensitivity to its model parameters. A further analogy can be taken from support vector machines, which commonly apply transformations to the space of the dataset to arrive in a final space which specifically amplifies the differences between the global features of the data.

### b. Information Theory

As we have discussed, the unreliability of the optimization can be attributed to the flatness of the $l^2$ norm, which itself can be attributed to the insensitivity of the model behavior with respect to $\boldsymbol{\theta}$. One might then question the case when the $l^2$ norm is completely flat for all values of $\boldsymbol{\theta}$. In this case, the parameters of the model would not control the description of the data as different values of $\boldsymbol{\theta}$ would describe the same phenomena with equal validity. In other words, the information provided by the data would be completely useless in identifying the optimal values of $\boldsymbol{\theta}$. In most cases the challenges may not be as drastic, but nevertheless, a robust procedure is necessary to ensure a proper fit for this particular parameter inversion problem. To fully quantify the information provided by a dataset about $\boldsymbol{\theta}$, we turn to information theory.

Here, we will use the log likelihood and the D-optimality criterion to assess how experimental conditions impact the information gathered about the optimal $\boldsymbol{\theta}$ of a material to then better design force-indentation experiments that would allow the user to obtain more meaningful data. While we will now present an informal description and motivation for the use of these two metrics, further details can be found in the technical appendix and more rigorous derivations and formal motivations can be found in the relevant literature[45,49–54].

First, consider representing the probability $P(\tilde{\sigma}_{obs}(\tilde{t}_0)|\boldsymbol{\theta})$ of correctly fitting the time domain model $\tilde{\sigma}$ at a single instant in time $\tilde{t}_0$ to a measured force-indentation curve $\tilde{\sigma}_{obs}$. As the noise in $\tilde{\sigma}_{obs}$ is assumed to follow a Gaussian process with standard deviation $s$, $P(\tilde{\sigma}_{obs}(\tilde{t}_0)|\boldsymbol{\theta})$ can then also be given as a Gaussian distribution as seen in Eqn. 6. Then, assuming the individual probability distributions for fitting each instant in time are independent and identically distributed, the probability (or likelihood) of correctly fitting the entire force-curve can be given as the product of the individual probability distributions for every instant in time. As the quality of the model's description of the data increases, so does the likelihood of proper fitting. Coincidentally, minimizing the $l^2$ norm with respect to $\boldsymbol{\theta}$ is equivalent to maximizing the likelihood with respect to $\boldsymbol{\theta}$. Hence the optimal parameter set is often referred to as the maximum likelihood estimate (MLE) of $\boldsymbol{\theta}$[55,56]. For the sake of convenience, the logarithm of the likelihood product is used, thus resulting in the log likelihood $L(\tilde{\sigma}_{obs}|\boldsymbol{\theta})$ (Eqn. 7 and 7a).

$$P(\tilde{\sigma}_{obs}(\tilde{t}_0)|\boldsymbol{\theta}) = \frac{1}{\sqrt{2\pi s^2}} e^{-\frac{(\tilde{\sigma}(\tilde{t}_0,\boldsymbol{\theta})-\tilde{\sigma}_{obs}(\tilde{t}_0))^2}{2s^2}} \tag{6}$$

$$L(\tilde{\sigma}_{obs}|\boldsymbol{\theta}) = \sum_{\tilde{t}} -\frac{\log(2\pi s^2)}{2} - \frac{(\tilde{\sigma}(\tilde{t},\boldsymbol{\theta})-\tilde{\sigma}_{obs}(\tilde{t}))^2}{2s^2} \tag{7}$$

$$L(\tilde{Q}_{obs}|\boldsymbol{\theta}) = \sum_{\omega} -\frac{\log(2\pi s^2)}{2} - \frac{(\tilde{Q}(\omega,\boldsymbol{\theta})-\tilde{Q}_{obs}(\omega))^2}{2s^2} \tag{7a}$$

$$J(\boldsymbol{\theta}_{MLE})_{i,j} = \left[-\partial_{\theta_i}\partial_{\theta_j}L(\tilde{\sigma}_{obs}|\boldsymbol{\theta})\right]_{\boldsymbol{\theta}=\boldsymbol{\theta}_{MLE}} \quad (8)$$

$$J(\boldsymbol{\theta}_{MLE})_{i,j} = \left[-\partial_{\theta_i}\partial_{\theta_j}L(\tilde{Q}_{obs}|\boldsymbol{\theta})\right]_{\boldsymbol{\theta}=\boldsymbol{\theta}_{MLE}} \quad (8a)$$

As previously stated, the flatness, or curvature of the $l^2$ norm allows us to build an insight into how to quantify the information contained within a dataset. Again, if the $l^2$ norm surface is entirely flat, changes in $\boldsymbol{\theta}$ will not alter the model's description of the data – therefore, the data provides no information about the optimal value of $\boldsymbol{\theta}$, given as $\boldsymbol{\theta}_{MLE}$. Conversely, if the $l^2$ norm, or equivalently the log likelihood, is highly curved in the vicinity of $\boldsymbol{\theta}_{MLE}$, then, the dataset is considered to contain a large amount of information about the $\boldsymbol{\theta}_{MLE}$. Thus, the curvature of the log likelihood in the vicinity of $\boldsymbol{\theta}_{MLE}$ seems to be a sensible metric to quantify the information that the dataset contains about $\boldsymbol{\theta}_{MLE}$. Indeed, the matrix of mixed second partial derivatives $J(\boldsymbol{\theta}_{MLE})_{i,j}$ of the log likelihood, evaluated at $\boldsymbol{\theta}_{MLE}$ (Eqn. 8 and 8a) is a commonly used measurement to quantify this information and is referred to as the Observed Fisher Information matrix. Various elements of $J(\boldsymbol{\theta}_{MLE})_{i,j}$ can be used to quantify the information contained in a dataset[49,50]. Here, we consider the determinant $\det(J)$ of the matrix, referred to as the D-optimality criterion. Of course, while we provided the motivation and informal derivations of $\det(J)$ and $L(\tilde{\sigma}_{obs}|\boldsymbol{\theta})$ using the example of force-indentation curves (time domain), the results are equally valid for the frequency domain model of the modulus.

### i. Time Domain

First, we consider the behavior of the information $\det(J)$ of a force-indentation experiment on a material with a single relaxation time $\tilde{\tau}_1$ (SLS, Fig. 1a) in the time domain. In Fig. 4a, the distribution of $\det(J)$ increases exponentially as a function of the dimensionless time, obtaining its maximum value at the end of the experiment (a longer experiment provides more information than a shorter one, generally speaking, since it allows greater probing of the material behaviour, assuming a uniform strain rate). While the relaxation time is not mathematically restricted to be shorter than the length of the experiment, it is evident that the information available for larger relaxation times is much diminished in comparison to those relaxation times that are shorter. This is as expected, since such relaxation times would not be reached in shorter experiments, and therefore, their effects could not be observed. Most noteworthy is the behavior of the information after $\tilde{\tau}_1$ exceeds $T/10$ as evidenced by the exponential decrease in the information calculated at the end of the experiment, seen in Fig. 4b. We can then deduce that relaxation times larger than $T/10$ do not contribute significantly to the behavior of the model (within an experiment of length $T$), a finding that is consistent with the flatness seen in the regions of the contour maps of the $l_t^2$ norm for values of $\tilde{\tau} > 0.1$. A similar result was previously reported for stress-relaxation experiments wherein the information obtained about relaxation times exceeding $T/5$ was significantly diminished[45]. As the nondimensionalization of the timescales with respect to the experiment length is equivalent to normalizing by the inverse of the strain rate, it is evident that this upper bound can also be expressed as $1/10\ \varepsilon_0$. From here, we can see that exciting the material with a more abrupt strain input results in a more restricted upper bound as the experiment probes the shorter timescales of the material. By straining the material very slowly, the upper bound increases, allowing one to access the response of the material at larger timescales.

Conversely, as the relaxation time decreases, the information increases, indicating that shorter relaxation times will be easier to identify. However, as $\tilde{\tau}$ approaches the sampling timestep $\Delta t$ of the experiment, this information gain stagnates. In this case, the relaxation process governed by $e^{-\tilde{t}/\tilde{\tau}_n}$ is not adequately sampled in the experiment and thus, relaxation processes that occur in a period shorter than the sampling frequency do not lead to a change in the behavior of the model.

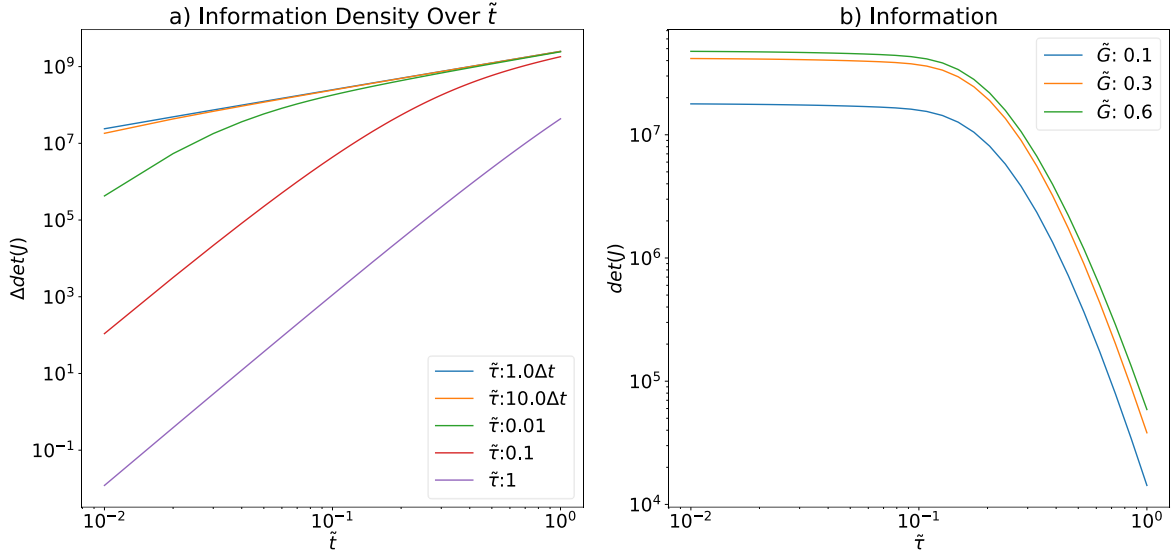

**Figure 4**: a) Distribution of information over the dimensionless time for SLS materials with various relaxation times. b) The behavior of the information as a function of the relaxation time for various moduli.

A further restriction can be placed on the lower bound of the relaxation time when the general convolution definition for the stress-strain or force-indentation relationship is used, such as in Eqn. 1 and 1a, rather than assuming a linearized excitation. In this case, as the data from the experiment is a discrete signal, the convolution operator is inherently discretized. However, approximating continuous convolutions like those in Eqn. 1 and 1a with a discretized convolution is only valid when the kernel $Q$ and the input $\varepsilon$ or $h^\beta$ are band limited functions[56]. Of course, $Q$ is not generally band limited; however, as shown in the technical appendix, $Q$ can be approximately treated as being band limited only when the shortest relaxation time is several orders of magnitude greater than the sampling timestep of the experiment:

$$\tau_{min} > 1000 \, \Delta t \qquad (9)$$

To demonstrate how these restrictions impact the behavior of materials with multiple relaxation times, we first simulate the force-indentation behavior of a material with two relaxation times $\tilde{\tau}_1, \tilde{\tau}_2$ and two moduli $\tilde{G}_1, \tilde{G}_2$. We then assess the description of this data by a model with a single relaxation time (SLS) by calculating the normalized maximum log likelihood $L(\tilde{\sigma}_{obs}|\boldsymbol{\theta}_{MLE})$. In Fig. 5a and d, we plot $L(\tilde{\sigma}_{obs}|\boldsymbol{\theta}_{MLE})$ for various combinations of $\tilde{\tau}_1, \tilde{\tau}_2$ and $\tilde{G}_1, \tilde{G}_2$. As the difference between the two relaxation times $\tilde{\tau}_1, \tilde{\tau}_2$ of the material increases, the ability of the SLS model in describing the features of the data diminishes as evidenced by the decrease in $L(\tilde{\sigma}_{obs}|\boldsymbol{\theta}_{MLE})$. Of course, as $\tilde{\tau}_1$ and $\tilde{\tau}_2$ approach each other, the opposite is true as their behaviors

can be equivalently treated as a single relaxation time. As seen in the wide distributions in $L(\tilde{\sigma}_{obs}|\boldsymbol{\theta}_{MLE})$, if the difference between $\tilde{\tau}_1$ and $\tilde{\tau}_2$ drops within 1-2 orders of magnitude, the likelihood that the data can be described by a single relaxation time is maximized. In such a case, the two nearly coincidental relaxation times would not be resolved from the analysis of the experimental data in the time domain. For example, consider the blue curve in Fig. 5a. Here, $\tilde{G}_1 = \tilde{G}_2 = 0.5$, $\tilde{\tau}_1 = 0.001$s, and various values of $\tilde{\tau}_2$ are given along the x-axis of the figure. In this case, only values of $\tilde{\tau}_2 > 0.1$s can be distinguished in the time domain (i.e., only for values of $\tilde{\tau}_2 > 0.1$s is it possible to determine that having a single relaxation time in the model is insufficient to fit the data). To better illustrate this point, the data and the SLS fit associated with $\tilde{\tau}_2 = 0.01$s is provided in Fig. 5c. As expected from the preceding discussion, the two relaxation times cannot be determined, thus the SLS model adequately and more concisely describes the material.

A similar phenomenon occurs as one of the relaxation times approaches or surpasses the length of the experiment, or the inverse of the strain rate. Even if the two relaxation times are separated by several orders of magnitude (for instance $\tilde{\tau}_1 = 0.001, \tilde{\tau}_2 = 10$ as seen in the blue curve in Fig. 5a), the slower relaxation process will not be able to have a significant effect on the behaviour of the material during the experiment and thus, will not be detected in the experiment (since $\tilde{\tau}_2 = 10$ is ten times longer than the duration of the experiment). This explains why the curves in Fig. 5a and 5d rise after reaching a minimum when $\tilde{\tau}_2$ is approximately unity).

A further example can be seen in the orange log likelihood curve in Fig. 5a and the corresponding set of force curves seen in Fig. 5b. In this example, the relaxation times are separated by nearly three orders of magnitude; therefore, the insufficiency using single relaxation time in the model is readily identified in the time domain and one can easily conclude that SLS description of the data is inaccurate. An additional example is provided in Fig. 5d for the case when one of the moduli of the material is small compared to its total stiffness. Here, similar features of $L(\tilde{\sigma}_{obs}|\boldsymbol{\theta}_{MLE})$ are seen wherein the behaviors of the relaxation times are most noticeably different when they are separated by several orders of magnitude. However, as one of the moduli is smaller than for the previous example, the effects associated with this modulus are similarly negligible. Therefore, the magnitude of $L(\tilde{\sigma}_{obs}|\boldsymbol{\theta}_{MLE})$ for the SLS is significantly increased for most of the horizontal axis range in Fig. 5d when compared to Fig. 5a, thus suggesting that relaxation times should only be introduced when they are associated with moduli that are roughly equal in magnitude.

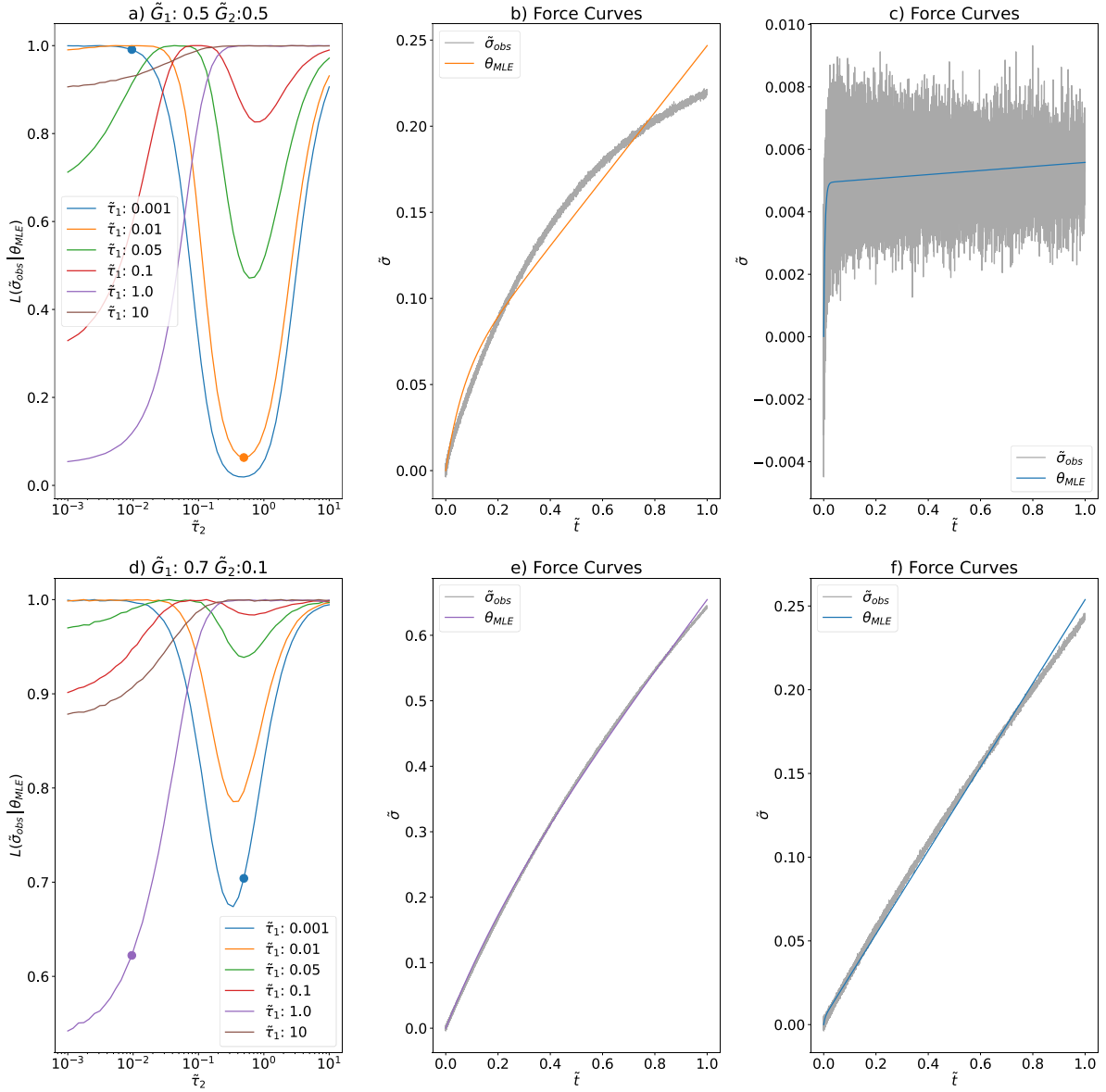

**Figure 5**: The maximum log likelihood $L(\tilde{\sigma}_{obs}|\boldsymbol{\theta}_{MLE})$ for an SLS description of a material with two relaxation times $\tilde{\tau}_1, \tilde{\tau}_2$ and $\tilde{G}_1, \tilde{G}_2$. a) shows $L(\tilde{\sigma}_{obs}|\boldsymbol{\theta}_{MLE})$ normalized to a 0-1 scale for a material given by $\tilde{G}_1 = \tilde{G}_2 = 0.5$ with various combinations of relaxation times. b) shows the force-indentation curve corresponding to the circular orange marker in Fig. 5a and the most likely SLS description. c) shows a similar correspondence for the circular blue marker in Fig. 5a. d) shows values of $L(\tilde{\sigma}_{obs}|\boldsymbol{\theta}_{MLE})$ normalized to a 0-1 scale for a material given by $\tilde{G}_1 = 0.7$ and $\tilde{G}_2 = 0.1$ ($\tilde{G}_1$ noticeably larger than $\tilde{G}_2$) for various relaxation times. e) shows a correspondence for the circular purple marker in Fig. 5d. f) shows a correspondence for the circular blue marker in Fig. 5d.

We can then conclude that for force-indentation curves in the time domain, relaxation times can most reliably be extracted when their values fall between $\Delta t$ and $T/10$ (or equivalently $1/10\ \varepsilon_0$), when using a linearized force-indentation model as we have done in our example. If the force-indentation relation is kept general (i.e., without specifying its form) and the (discretized) convolution definition is used, then the lower bound increases to about $1000\,\Delta t$. Furthermore, the presence of multiple relaxation times can only be determined if they are separated by at least 1-2 orders of magnitude and within the previously mentioned ranges. Additionally, the magnitude of the moduli of each Maxwell arm should be comparable and larger than the amplitude of the measurement noise. Similarly, as the magnitude of both the log likelihood and the information scale inversely with the square of the amplitude of the noise, the ability to determine these parameters drastically diminishes as the amount of noise increases. Violating these criteria introduces the possibility of fitting these parameters to artifacts unrelated to the material.

### ii. Frequency Domain

A similar approach can be taken to determine the range of detectable relaxation times in the frequency domain. As was done in the time domain, we first assess the shape of the information distribution $\det(J)$ of a model with a single relaxation time as a function of frequency as seen in Fig. 6a. For relaxation times between the sampling timestep and the length of the experiment, the information increases until the frequency passes the relaxation frequency $\widetilde{\omega}_r = 1/\widetilde{\tau}$. After this point, the information slowly declines with frequency; however, this feature is an artifact of the normalization of $\widetilde{Q}$ with respect to $G_g$ since now all the models share a common behavior for frequencies greater than $1/\widetilde{\tau}$.

The behavior of the information in $\widetilde{Q}$ can be understood by investigating the behavior of the model as a function of frequency. Fig. 7 shows the storage and loss moduli of $\widetilde{Q}$ for various values of $\widetilde{\tau}$. The behavior of the storage modulus follows a sigmoidal-like transition between $G_e$ and $G_g$ whereas the loss modulus follows a shape similar to a Boltzmann distribution. As the locations of these features are controlled by $\widetilde{\omega}_r$, increasing $\widetilde{\tau}$ past the resolution of the frequency axis will cause these features to not be adequately sampled during the experiment as seen in the pink curves in Fig. 7a and b. As the resolution of the frequency axis is given as $T$, relaxation times that are longer than the length of the experiment will not be properly sampled. Therefore, we can understand the decaying behavior of the information for large values of $\widetilde{\tau}$ as seen in Fig. 6b. Of course, this behavior is consistent with what is seen in the time domain; however, the decay is not as abrupt in the frequency domain compared with the time domain, as seen in Fig. 4b.

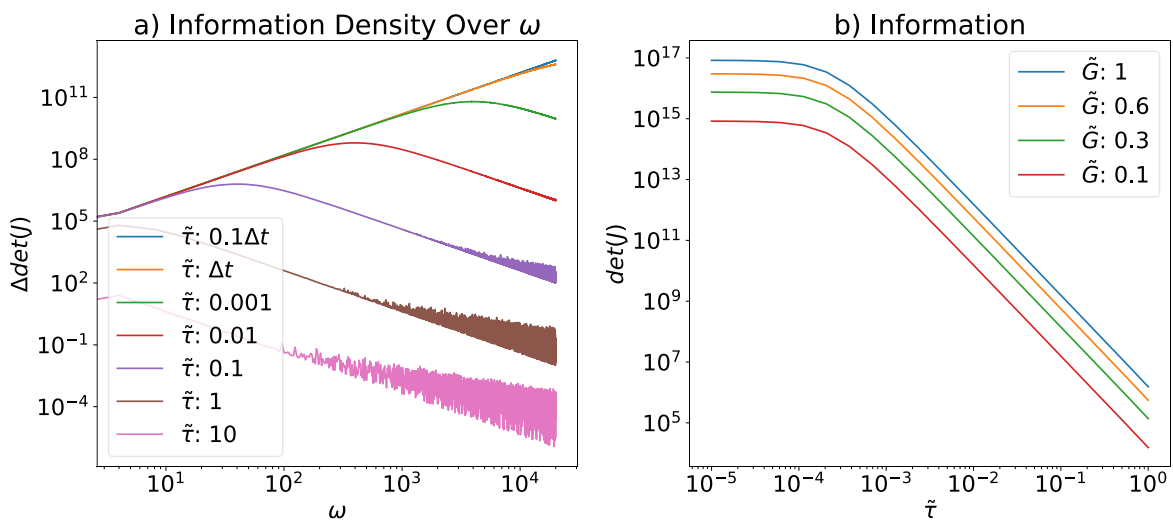

**Figure 6**: a) Distribution of information over the dimensionless time for SLS materials with various relaxation times with $\tilde{G} = 1$ (fluid-like materials). b) Behavior of the information as a function of the relaxation time for various moduli.

On the opposite end, as the relaxation time decreases, the magnitude of the information increases until it meets a stagnation point as also seen in the time domain. In this case, the location of the previously mentioned features of the storage and loss moduli shifts outside of the Nyquist band when $\tilde{\tau}$ decreases past $2\,\Delta t$. Therefore, relaxation times that happen faster than the length of the sampling window will not be obtained in the frequency domain. It is interesting to note that the model appears to correspond to two different 'elastic' solids when $\tilde{\tau}$ is outside of these detectable bounds. On one end, the material relaxes so quickly as to appear instantaneous – on the other, the material relaxes so slowly as to appear as if it doesn't relax at all. Both cases result in approximately flat storage and loss moduli with the flatness increasing as $\tilde{\tau}$ approaches either 0 or $\infty$[32].

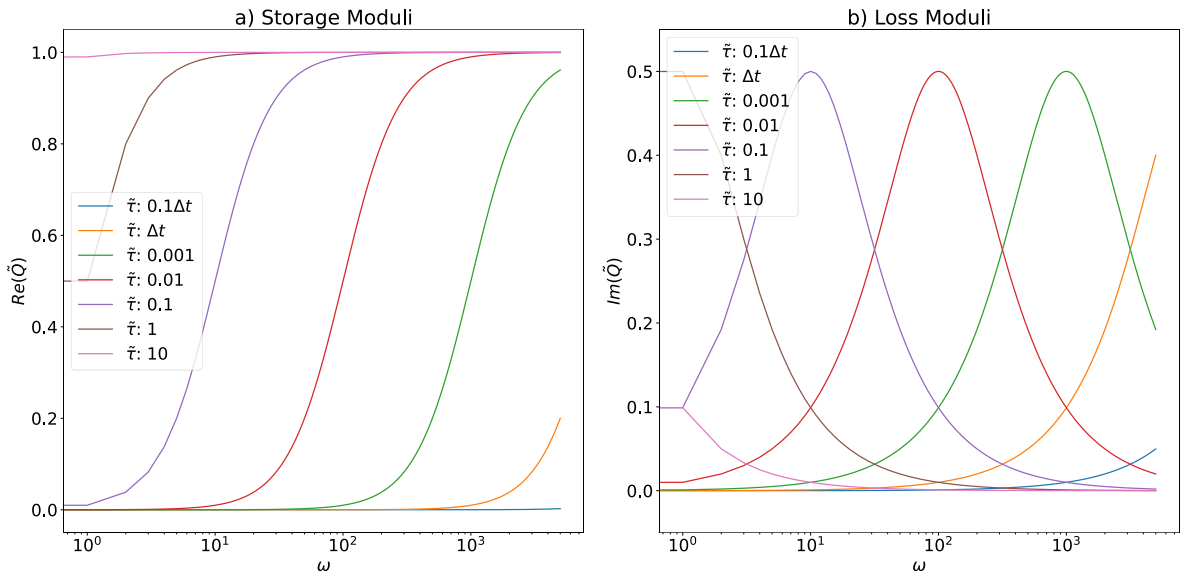

**Figure 7**: Behavior of the a) storage and b) loss moduli of the SLS model for various relaxation times with $\tilde{G} = 1$ (fluid-like materials). As stated in the text, the storage modulus behaves as a sigmoid, with its transition period being governed by the inverse of the relaxation time. The loss modulus follows a similar shape as a Boltzmann distribution, with its maximum value occurring at the inverse of the relaxation time.

It should be noted that $2\,\Delta t$ serves as an idealized lower limit for the detectable relaxation times. In practice, noise may obscure the behavior of the material at high frequencies. When deriving the model for $\tilde{Q}$ that was used in obtaining these results, we used the explicit form of $\tilde{Q}$ for a generalized Maxwell model, therefore avoiding taking the modified Fourier transform of the force-indentation curves (i.e., $\tilde{Q}$ was not obtained from data, but instead from its analytical expression). We further assumed that the noise would be linearly additive to $\tilde{Q}$; however, when calculating $\tilde{Q}$ directly by inverting the force-indentation data, the noise does not follow a simple linear behavior. In the technical appendix we demonstrate that the expected value of the magnitude of the noise induced error in $\tilde{Q}$ scales with the square of the amplitude of the noise and inversely with the square of the magnitude of the transformed strain as seen in Eqn. 10. As the magnitude of the modified discrete Fourier transform of monotonically increasing linear-like inputs (as are typical in force-indentation experiments – further demonstrated in the appendix) rapidly diminishes at high frequencies, high frequency error is commonly observed when calculating $\widetilde{Q}$. This high frequency error can be thought of as noise with an amplitude, or standard deviation, which increases as a function of frequency. As mentioned before, since the information $\det(J)$ scales inversely with the square of the amplitude of the noise, we can expect that the increasing noise will result in a more rapid attenuation of the information at high frequencies and thus, it will impose greater restrictions on the lower limit of detectable relaxation times. To avoid this issue, one can use strain excitations that decay more slowly in the modified Fourier domain. Since these types of 'weakly decaying' spectra correspond to more slowly diverging excitations in the time domain, it is advantageous from a noise perspective to perform slower experiments. According to the results established in the time domain section, this also corresponds to decreasing the strain rate, thus increasing the upper bound of accessible relaxation times. Alternatively, a slower diverging signal has a wider bandwidth in the modified Fourier domain, thus allowing for a broader, more reliable characterization of the material response.

$$\delta_Q \sim \frac{s^2}{|\varepsilon(\omega)|^2} \tag{10}$$

As we did in the time domain, we will now assess the behavior of materials with multiple relaxation times in the frequency domain. We will again investigate the maximum log likelihood $L(\widetilde{Q}_{obs}|\boldsymbol{\theta}_{MLE})$ of an SLS model in describing a material with two relaxation times $\tilde{\tau}_1, \tilde{\tau}_2$. As seen in the same calculation done for the time domain, the SLS model can describe such a material only when the difference between $\tilde{\tau}_1$ and $\tilde{\tau}_2$ is small. However, unlike the time domain, the differences in $\tilde{\tau}_1$ and $\tilde{\tau}_2$ for which this single relaxation time approximation is valid are much smaller, indicating that one can more easily discern when additional relaxation times are needed in the model. In Fig. 8a and d, the peak-like distributions formed in $L(\widetilde{Q}_{obs}|\boldsymbol{\theta}_{MLE})$ indicate regions where the single relaxation time approximation is valid. Compared to the corresponding plots in the time domain seen in Fig. 5a and d, the distributions formed here are much narrower and larger in magnitude, again suggesting that the effects associated with the relaxation times are more pronounced in the frequency domain compared to the time domain. Such a behavior should be expected, however, as the number of distinct relaxation times is directly linked to the number of distinct peaks in the loss modulus.

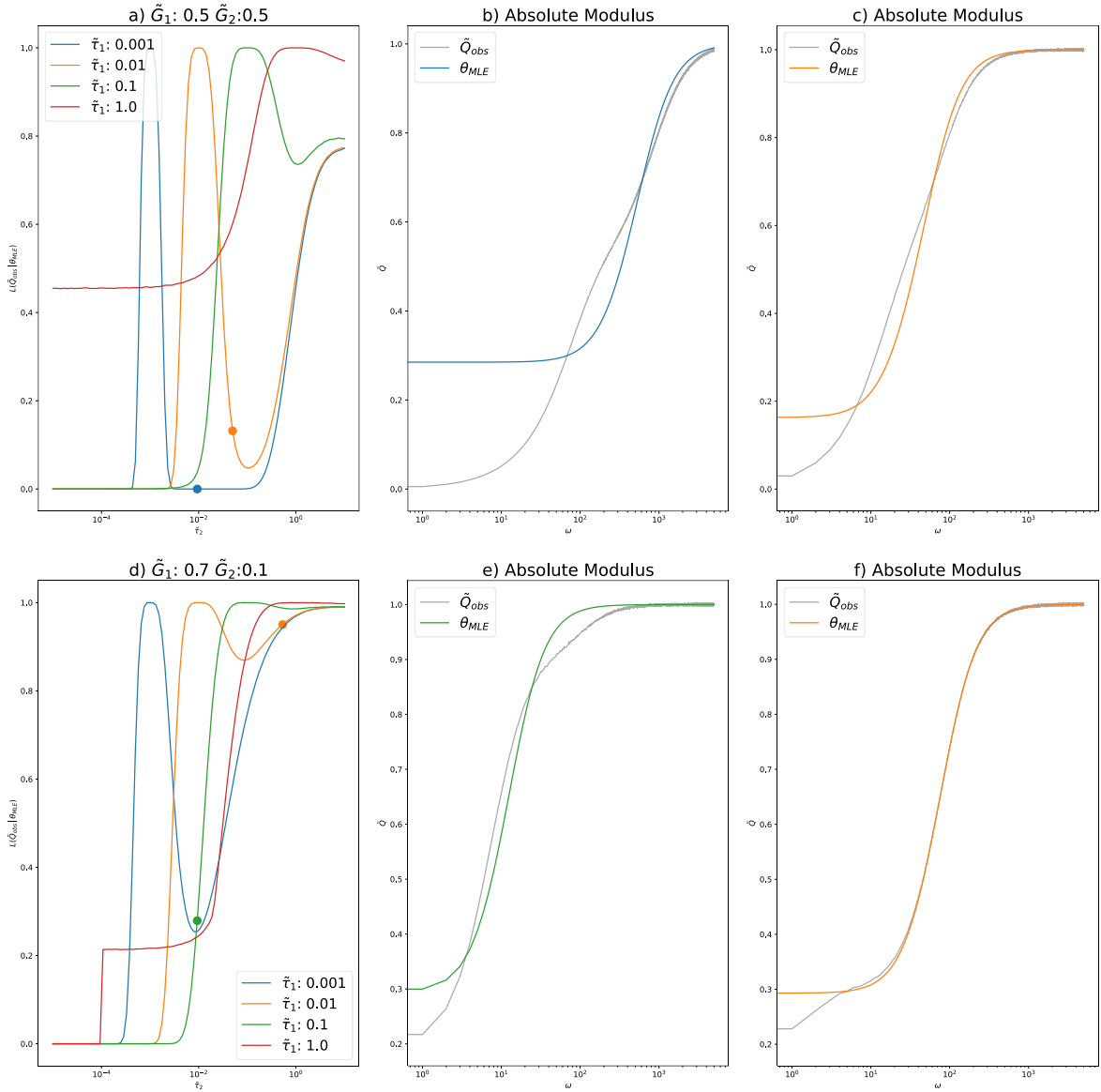

**Figure 8**: The maximum log likelihood $L(\widetilde{Q}_{obs}|\boldsymbol{\theta}_{MLE})$ for an SLS description of a material with two relaxation times $\tilde{\tau}_1, \tilde{\tau}_2$ and $\tilde{G}_1, \tilde{G}_2$. a) shows $L(\widetilde{Q}_{obs}|\boldsymbol{\theta}_{MLE})$ normalized to a 0-1 scale for a material given by $\tilde{G}_1 = \tilde{G}_2 = 0.5$ with various combinations of relaxation times. b) shows a correspondence between the circular blue marker in Fig. 8a and its associated absolute moduli for the material and the most likely SLS description. c) shows a similar correspondence between the circular orange marker in Fig. 8a. d) shows values of $L(\widetilde{Q}_{obs}|\boldsymbol{\theta}_{MLE})$ normalized to a 0-1 scale for a material given by $\tilde{G}_1 = 0.7$ and $\tilde{G}_2 = 0.1$ for various relaxation times. e) shows a correspondence for the circular green marker in Fig. 8d. f) shows a correspondence for the circular orange marker in Fig. 8d.

As was done in Fig. 5, color-coded markers have been placed on top of the plots in Fig. 8a and d to correspond to individual plots of the behavior of the SLS description of various two-relaxation-time materials in the frequency domain. Consider first the blue marker in Fig. 5a: as the log likelihood is significantly reduced, we can imagine that the SLS model should not be able to describe the material. Indeed, this is evident in Fig. 5b as the error in $\widetilde{Q}$ propagates across a few decades of frequency. Such an example illustrates very well the distinction between the frequency and time domains as here, we see that materials with relaxation times within an order of magnitude can be resolved whereas the time domain required 1-2 orders of magnitude of separation. Furthermore, we see that the magnitude of the normalized log likelihood approaches a value of 0 for a significant range of separation between the relaxation times. Such a feature was not present in the corresponding time domain plot. Further examples can be seen in Fig. 8c, e, and f.

We can thus observe that the frequency domain offers a significant advantage over the time domain in determining the parameters of generalized Maxwell models. Specifically, relaxation times between $2\,\Delta t$ and $T$ can be individually determined, with the behaviour of multiple relaxation times being able to be resolved if they are separated by more than an order of magnitude. The results have been summarized in Table 1 below.

**Table 1**: Summarized description of the minimum and maximum relaxation times obtained from the time and frequency domain analysis of force-indentation experiments as well as the minimum resolution for which multiple relaxation times can be determined.

|  | $\tau_{min}$ | $\tau_{max}$ | $\Delta\tau_{min}$ |
|---|---|---|---|
| **Time domain** | $\Delta t^*$ | $T/10$ | 100x |
| **Frequency domain** | $2\,\Delta t^*$ | $T$ | 10x |

*Note that the lower bound for the time domain will increase by 3 orders of magnitude if the discrete convolution is used to calculate the force-indentation model. Additionally, the lower bound of the frequency domain will also increase in the presence of noise in the direct inversion of the force-indentation data.*

4. **Demonstration with Experiments**

We can now demonstrate how these criteria can be leveraged to better design force-indentation (FD) experiments to more accurately characterize a polydimethylsiloxane (PDMS) sample. In this demonstration, we will use AFM FD curves as well as uniaxial compression tests done on the macro-scale with a Universal Testing Machine (MTS) (see Methods section below). As the compression tests are performed by applying a linear strain input, the results of the previous sections should also apply. As a reminder, we have determined that in the time domain, one can only obtain relaxation times between the sampling timestep $\Delta t$ and one tenth of the length of the experiment $T/10$. Within these bounds, multiple relaxation times can only be determined if they are separated by 2 decades of time and if they contribute similar values of their respective moduli. Alternatively, if one opts to use the frequency domain, the maximum range of detectable relaxation times increases to be between $2\Delta t$ and $T$ and the required minimum spacing between multiple relaxation times decreases to a single decade.

Thus, to design an AFM experiment which maximizes the information obtained about the characteristic timescales of PDMS, we will first have to estimate the relaxation time(s) for the material using commonly accepted data. As the relaxation time of a material is roughly given as the ratio of its viscosity to its elasticity, we estimate the relaxation time of PDMS by dividing accepted values for its viscosity by its accepted modulus, given as $1\,kPa\,s$ and $1\,MPa$, respectively[57,58]. The resulting estimate for $\tau$ is then found to be $0.001\,s$. Therefore, we specify our AFM experiment to have a sampling timestep of $2e-5\,s$ (the lower limit on this particular AFM instrument) and an experiment length of approximately $0.1s$. Additionally, the approach velocity was specified to give a strain rate of roughly $40\frac{\%}{s}$ (indentation normalized by probe radius). Thus, the boundaries of accessible timescales from the AFM experiment are determined to be at most $2e-5\ s$ and $0.01s$ in the time domain, and at most $4e-5\ s$ and $0.1s$ in the frequency domain. However, it should be noted that if we use the general convolution definition for fitting the force-indentation curves in the time domain, the lower limit increases to only $0.02s$, about an order of magnitude larger than the characteristic timescale of PDMS. Additionally, these restrictions on the time domain analysis limit us to only characterize a single relaxation time. To utilize the full range of information available from the AFM experiment, one would have to assume a functional form for the indentation, which is hardly ideal, or simply use the frequency domain approach. To further extend the range of the timescales that can be obtained from this experiment, one would need to decrease the sampling rate and increase the length of each experiment. In the frequency domain, however, we can obtain $\tau$ values between $4e-5\ s$ and $0.1s$. Furthermore, since we can capture the behavior of multiple relaxation times separated by only a single decade, we can obtain a much greater resolution than compared to the time domain. For example, an experiment length of $0.1\ s$ allows the determination of, at most, four unique relaxation times – far greater than the single relaxation time obtained in the time domain.

The macroscale mechanical tester used here is typically applied in much slower experiments as compared to the AFM, applying a strain rate of only $0.2\frac{\%}{s}$. As a result of the slower deformation rate, the sampling timestep was restricted to only $0.1s$ and the length of the experiment was around $10s$. Thus, we can expect that the macroscale experiment will yield a superb characterization of the low frequency values of $Q$ (longer timescale relaxations) but will be incapable of describing the expected $0.001\ s$ relaxation time of the material.

Fig. 9a and b show the spectral averaged storage and loss moduli of the material obtained from direct modified Fourier inversion of the AFM force-indentation and MTS stress-strain curves corresponding to our established method[42]. The set of force-indentation and stress-strain curves can be seen in Fig. 9c and d. The values of the storage and loss moduli (dotted scatter plots) obtained from the two devices agree exceptionally well within the small frequency range of overlap. Furthermore, both of these results agree quite well with accepted values for the storage and loss modulus of PDMS[59]. As the Nyquist frequency limit of the macroscale MTS experiment is restricted to only 5 Hz, the observed behaviour of the PDMS is mostly elastic – having a flat storage modulus and negligible loss modulus. This feature was consistent when fitting both the time and frequency domain constitutive equations for the Maxwell model (Eqn. 3 and 3a) to the respective time and frequency domain data as can be seen in Table 2. Here, the Maxwell model parameters obtained from the fitting procedures indicate that PDMS is an elastic material with a modulus of $\sim0.5 MPa$ – in relatively good agreement with the accepted $\sim0.9\ MPa$. Of course, we know this conclusion is false; however, the MTS experiment was not able to probe the relevant timescales of this viscoelastic behavior due to its limited sampling timestep.

As existing dynamic mechanical analyses of PDMS have been limited to only a few hundred Hz, the plateauing behaviour of the storage modulus, a hallmark of Maxwell materials, is usually not fully captured as it is in the AFM data in Fig. 9a. In the presence of such restrictions on the acquisition of high frequency data, practitioners typically resort to using power law models which fit the data well in the low frequency regime, but fail to capture the higher frequency plateau, if present. Like the MTS data, we fit Maxwell models (Eqn. 3 and 3a) to the time and frequency domain data obtained from the AFM experiments. As a reminder, we had specified the sampling rate, experiment length, and linearized strain rate such that the estimated relaxation time of the PDMS falls within these boundaries. In the time domain, however, we chose to use the general convolution definition rather than assume a functional form for the indentation which resulted in a lower bound which was greater than this estimated time constant. As a result, the quantities identified from the time domain, although they agree with the force curves, do not agree with the frequency domain data. As we have come to expect, the frequency domain provides quantities which agree with the behavior of the material in both the time and frequency domains.

To further demonstrate the quality of this agreement, we have supplied the mean squared error between fitted force-indentation and stress-strain curves using the parameters obtained from both the time and frequency domain approaches with the AFM and MTS. As seen in these errors as well as in Fig. 9c and d, the qualitative behaviour of these curves agrees quite well, with the frequency domain tending to provide a slightly more optimal result than the time domain (note that the force curve for the frequency domain parameters was constructed using a high order polynomial fit to the data as the general convolution definition would not work for the relaxation time obtained from this parameter set). We can then conclude that due to the larger range of detectable timescales available in the spectral analysis of the AFM experiments, we are able to detect the viscoelastic nature of the material. Since the MTS experiment does not adequately interrogate the timescale of the rate-dependent behaviour of the PDMS, we are lead to incorrectly conclude that the material is elastic.

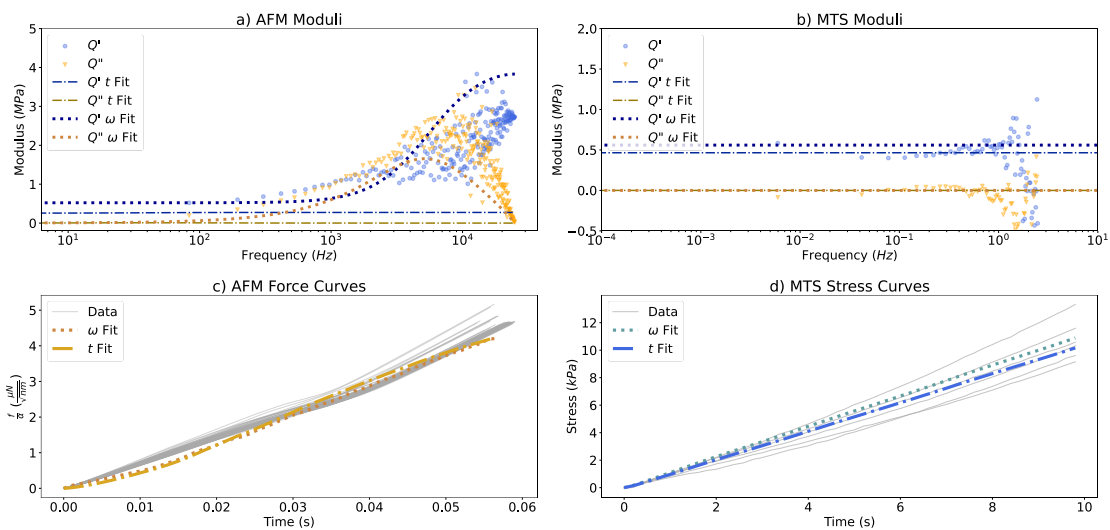

**Figure 9:** a) the spectral averaged storage (blue scatter) and loss (yellow scatter) moduli obtained from the direct inversion of the AFM force-indentation curves. b) The spectral averaged storage (blue scatter) and loss (yellow scatter) moduli obtained from the direct inversion of the MTS stress-strain curves. c) The set of force-indentation curves from the AFM experiment (grey) and corresponding time (dashes) and frequency (squares) domain fits. d) The set of stress-strain curves from the MTS experiment (grey) and corresponding time (dashes) and frequency (squares) domain fits.

**Table 2**: SLS parameters obtained from the characterization of PDMS using AFM and macroscale testing data in the time and frequency domain. We also include the ensemble averaged mean squared error between the fitted force-indentation and stress-strain curves (parameters obtained from fitting in both time and frequency domains) and the data obtained from the experiments.

|  | $G_e$ (MPa) | $G_1$ (MPa) | $\tau_1 (s)$ | MSE Time |
|---|---|---|---|---|
| **AFM** $t$ | 0.0 | 0.3 | 0.06 | $4e-14$ |
| **AFM** $\omega$ | 0.5 | 5.0 | 0.0002 | $2e-14$ |
| **MTS** $t$ | 0.6 | 0.0 | 0.02 | $9e6$ |
| **MTS** $\omega$ | 0.5 | 0.0 | 0.9 | $7e6$ |

## 5. Conclusions

Although AFM researchers have branched out into the fields of mechanobiology and micro-rheology, the instrument has yet to find wide-spread adoption within these communities and this work largely remains driven by AFM practitioners. In a sense, the AFM is a superior device in terms of its broad capabilities compared to colloid based rheology devices; however, there needs to be more focus on the reliability of the data obtained from AFM experiments. As we have shown, the range of timescales that can be obtained from force curves analysed in the time domain is at best limited to be less than a tenth of the length of the experiment and more than the sampling period. Furthermore, the use of the general convolution definition with discrete data may lead to severe inaccuracies which can only be avoided by increasing the lower bound by roughly a factor of 1000. To provide more context, such a finding is of significant importance to high throughput force volume measurements where individual force curves are performed in less than $1e-3s$ with a sampling period on the order of $1e-6s$. In these cases, the discretized convolution will never yield accurate results. Although we have demonstrated that performing this analysis in the frequency domain allows a much wider range of accessible timescales for the same experiment, we assumed that the noise, which stems from the time domain, would add linearly to the frequency domain. Such an assumption was necessary to simplify the analysis, but it is clear from the noise in the experimental data seen in Fig. 9, that this can cause serious issues. We further investigate the effects of idealized noise sources in the technical appendix, but a more rigorous analysis of the effects of more realistic noise sources would be beneficial to the community.

Additional guidance on the optimization of AFM FD experiments should consider alternative functional forms of the strain excitation. Here, we considered only linearized strain inputs which closely model real FD experiments, but can only be adjusted in terms of their strain rates. Specifically, impulse-like excitations may yield a broader interrogation of the timescales of the material. One could perhaps also express the problem as a functional optimization, seeking to obtain the strain input which maximizes an information functional. These types of advancements would allow the AFM to become a more robust rheological characterization tool.

## 6. Methods

A SYLGARD 184 Silicone Elastomer kit was used to prepare Polydimethylsiloxane (PDMS) following the standard procedure of mixing the base elastomer and the curing agent in a 10:1 mass ratio. The majority of the liquid was then poured into a flat-bottomed glass dish to a depth of 1 cm to be used for compression testing. The Atomic Force Microscopy (AFM) sample was made using 2 mL of the remaining liquid which was spin coated on a steel microscope sample disk at 100 rpm for 2 minutes. Both samples were then degassed and cured at room temperature for 24 hours.

A 10 mm biopsy punch was then used to extract 6 cylindrical samples from the glass dish. We then performed uniaxial unconfined compression testing (UUCT) with a Universal Testing Machine (MTS) (Applied Test Systems, Butler, PA, USA) on these samples using a 100 N load cell at a crosshead speed of 1 mm/min. The thickness of each sample was measured using a digital caliper with a sensitivity of 0.01 mm. The experimental stress was calculated as the applied force of the load cell divided by the cross-sectional area of the sample in direct contact with the UUCT device. Additionally, the strain was calculated as the ratio of the sample displacement to the original sample thickness.

AFM force spectroscopy measurements were then done in a room temperature environment filled with air with an MFP-3D (Asylum Research, Oxford Instruments, Santa Barbara, CA, USA) AFM using an ElectriCont-G (BudgetSensors, Izgrev, Sofia, Bulgaria) probe. The optical lever sensitivity was calibrated using the linear region of the repulsive part of a force curve performed on a freshly cleaned silicon wafer. The cantilever spring constant was determined to be 0.08617 N/m by thermal calibration using Sader's method[60]. 4 rounds of force spectroscopy experiments were performed on 4 different spots on the surface of the PDMS-coated steel disk. Each round of spectroscopy involved probing the surface of the sample 150 times within a confined region. Each force curve was specified with an approach velocity of 1.89 μm/s, a sampling frequency of 50 kHz, and a trigger force of 738.98 pN to stay within the small deformation criteria of contact mechanics. The probe was treated as a sphere with a radius of 25 nm as specified by the vendor information sheet.

**Acknowledgements**

M.R.M, K.S. and S.D.S. gratefully acknowledge support from the US National Science Foundation, under award CMMI-2019507. B.U. gratefully acknowledges support from the US Department of Energy, Office of Science, Basic Energy Sciences, under Award No. DE-SC0018041. M.S.A. and K.S. gratefully acknowledge support from the US National Science Foundation, under award CBET-2037849, and the ARCS Foundation Metro Washington Chapter.

**Technical Appendix**

1. $l^2$ **Norm**

Consider the case where a physical observable $f_{obs}$, such as the stress of a material, is measured in an experiment. This observable can then be modeled with a governing equation $f$ which takes a set of parameters $\boldsymbol{\theta}$ as its arguments. For example, the stress $\sigma_{obs}$ might be measured and described with an elastic governing equation which takes the Young's Modulus of the material $E$ as its argument, such as $\sigma(E)$. As the goal of the observation of $f_{obs}$ is to obtain the $\boldsymbol{\theta}$ which best explains the data (such as the best fitting Young's Modulus), we seek to minimize the error between the model prediction $f$ and the data $f_{obs}$. Such a problem can be solved as an optimization problem, where a distance metric is used to define the distance, or error, between the model prediction and the data. The most common distance metrics are the $l^1$ norm (Manhattan distance) and the $l^2$ norm (Euclidian distance, also referred to as the squared error). We use the $l^2$ norm as it is the most commonly used in these types of optimization problems in the AFM community. Also, the $l^2$ norm provides more general solutions, thus avoiding overfitting – although it can suffer performance issues when working with data with significant outliers[1]. As the $l^2$ norm sums over all the observations (typically instances of time) in the data, it can be thought of as a function of the parameter set $\boldsymbol{\theta}$.

$$l^2(\boldsymbol{\theta}) = \sum_n [f(n, \boldsymbol{\theta}) - f_{obs}(n)]^2 \qquad (S1)$$

2. **Log Likelihood**

In realistic conditions, the observed data $f_{obs}$ will have some experimental noise. Here, we treat the noise as a Gaussian process with a known standard deviation $s$. Then, for a single sample $n_0$ of the data, we can imagine the probability of correctly modeling $f_{obs}$ with the parameter set $\boldsymbol{\theta}$ which is given as $P(f_{obs}(n_0)|\boldsymbol{\theta})$. As the noise is generated by a Gaussian distribution, this probability will follow the same form. Then, as we previously sought to minimize the $l^2$ norm with respect to $\boldsymbol{\theta}$, we can now seek to maximize this probability, at least for the single sample $n_0$[1].

$$P(f_{obs}(n_0)|\boldsymbol{\theta}) = \frac{1}{\sqrt{2\pi s^2}} e^{-\frac{(f(n_0,\boldsymbol{\theta})-f_{obs}(n_0))^2}{2s^2}} \tag{S2}$$

Then, the probability of correctly modeling the entire dataset $P(f_{obs}|\boldsymbol{\theta})$ will be given as the joint probability distributions for every sample $n$. The resulting probability distribution can be significantly simplified if we assume that the distributions for each instant in time are independent and identically distributed (i.i.d.) which results in a product of the individual distributions[2–5].

$$P(f_{obs}|\boldsymbol{\theta}) = P(f_{obs}(1) \cap f_{obs}(2) \cap \ldots f_{obs}(N)|\boldsymbol{\theta}) = \prod_n^N P(f_{obs}(n)|\boldsymbol{\theta}) \tag{S3}$$

Again, this probability distribution governs the probability, or likelihood, that the model has correctly described the data. If the model perfectly describes the data, then this distribution will be maximized. In this case, it is highly *likely* that the model describes the data. Hence, this is commonly referred to as the likelihood. A far more commonly used quantity is the log likelihood $L(f_{obs}|\boldsymbol{\theta})$, which is given by the logarithm of the likelihood distribution. The product in the likelihood turns into a sum in the log likelihood, thus greatly simplifying the math when working with the quantity.

$$L(f_{obs}|\boldsymbol{\theta}) = \sum_n -\frac{\log(2\pi s^2)}{2} - \frac{\left(f(n,\boldsymbol{\theta}) - f_{obs}(n)\right)^2}{2s^2} \tag{S4}$$

Here, the connection between the $l^2$ norm and the log likelihood become obvious as seen in eqn. S5. We can now clearly see that minimizing the $l^2$ norm is equivalent to maximizing the log likelihood. As the logarithm is a monotonically increasing function, maximizing the log likelihood is then also equivalent to maximizing the likelihood. Thus, the $\boldsymbol{\theta}$ that either minimizes the $l^2$ norm or maximizes the likelihood / log likelihood is commonly referred to as the maximum likelihood estimate (MLE) $\boldsymbol{\theta}_{MLE}$.

$$L(f_{obs}|\boldsymbol{\theta}) = -\frac{l^2(\boldsymbol{\theta})}{2s^2} - \sum_n \frac{\log(2\pi s^2)}{2} \tag{S5}$$

### 3. Fisher Information, Observed Fisher Information, D-Optimality Criterion

Then, as we seek to perform experiments so that we obtain an accurate estimate of $\boldsymbol{\theta}_{MLE}$ for the physical observable $f_{obs}$, we might imagine that some data might be more useful than others. For example, if the data is heavily polluted with noise, determining $\boldsymbol{\theta}_{MLE}$ may be impossible. Thus, to measure the quality of a dataset for determining $\boldsymbol{\theta}_{MLE}$, we turn to information theory. Here, we will derive the Fisher Information and Observed Fisher Information using the results of the previous section.

To begin, we recall the hypothetical model $f$ for the data $f_{obs}$. First, consider the case where the model $f$ is entirely incapable of describing the features of $f_{obs}$. Here, one might have selected an entirely inappropriate model for the analysis! For instance, one could be attempting to model the data of the stress of an elastic material (with Young's Modulus $E$) under a constant strain input $\varepsilon(t \geq 0) = \varepsilon_0$ by using a viscous fluid model (with viscosity $\mu$ as the model parameter). Obviously in this ridiculous hypothetical case, the model and the data would never match, thus the log likelihood would be invariant under changes in the viscosity as seen in eqn.'s S6-S8. Furthermore, one could determine that the data does not contain any information about the viscosity of the material.

$$L(\sigma_{obs}|\mu) = \sum_n -\frac{\log(2\pi s^2)}{2} - \frac{\left(\sigma(n,\mu) - \sigma_{obs}(n)\right)^2}{2s^2} \tag{S6}$$

$$L(\sigma_{obs}|\mu) = \sum_n -\frac{\log(2\pi s^2)}{2} - \frac{(\mu\dot{\varepsilon}(n) - E\varepsilon(n))^2}{2s^2} \tag{S7}$$

$$L(\sigma_{obs}|\mu) = \sum_n -\frac{\log(2\pi s^2)}{2} - \frac{\varepsilon_0^2 E^2}{2s^2} \tag{S8}$$

As $L(f_{obs}|\boldsymbol{\theta})$ compares the model to the data, we see that if $\nabla_{\boldsymbol{\theta}} L(f_{obs}|\boldsymbol{\theta}) = 0$ (as in the previous example) then not only is the model incapable of describing the data, but also the data does not provide any information for determining $\boldsymbol{\theta}_{MLE}$. Conversely, if a change in $\boldsymbol{\theta}$ results in a significant change in $L(f_{obs}|\boldsymbol{\theta})$, then the data provides a similarly significant amount of information about $\boldsymbol{\theta}_{MLE}$. Thus, we see that the sensitivity of the log likelihood to changes in $\boldsymbol{\theta}$ serves as a way to measure the amount of information that the dataset contains about $\boldsymbol{\theta}$. We can mathematically define this sensitivity as the magnitude of the derivative of the log likelihood with respect to $\boldsymbol{\theta}$. This quantity is referred to as the Fisher Information $I(\boldsymbol{\theta})$ as seen in eqn. S9[2–6]. Here, the square is used to take the magnitude of the gradient. We could have just as well taken the absolute value; however, this would have resulted in a non-differentiable function. Finally, the expected value $E[\ ]$ is taken as the log likelihood is a probability distribution.

$$I(\boldsymbol{\theta}) = E\left[\left(\nabla_{\boldsymbol{\theta}} L(f_{obs}|\boldsymbol{\theta})\right)^2\right] \tag{S9}$$

Though straightforward, this definition is not practical as it explicitly depends on knowledge of the true parameter set, which is unknown. Hence, calculating the expected value is impossible. Instead, the Observed Fisher Information $J(\boldsymbol{\theta}_{MLE})$ can be used as an estimate for the true Fisher Information[2–4].

$$J(\boldsymbol{\theta}_{MLE})_{i,j} = \left[-\partial_{\theta_i}\partial_{\theta_j} L(f_{obs}|\boldsymbol{\theta})\right]_{\boldsymbol{\theta}=\boldsymbol{\theta}_{MLE}} \tag{S10}$$

Here, the Observed Fisher Information calculates the curvature of the log-likelihood with respect to the parameter set, evaluated at the maximum likelihood estimate $\boldsymbol{\theta}_{MLE}$. If the log likelihood is sharply defined at the maximum likelihood estimate, then changes in the $\boldsymbol{\theta}$ will result in significant changes in $L(f_{obs}|\boldsymbol{\theta})$, and hence, the Observed Fisher Information is large. In general, $J(\boldsymbol{\theta}_{MLE})_{i,j}$ defines the matrix elements for a matrix of size $m \times m$ where $m$ is the dimension of $\boldsymbol{\theta}$. Various elements of $J(\boldsymbol{\theta}_{MLE})_{i,j}$ can be used to quantify information, here we consider the D-optimality criterion, defined as the determinant of $J(\boldsymbol{\theta}_{MLE})$, for consistency with the study of the characteristics of step-like experiments[2,4]. While several technical details have been left out in obtaining $J(\boldsymbol{\theta}_{MLE})_{i,j}$ from $I(\boldsymbol{\theta})$, further details can be found elsewhere[4–8].

4. **Observed Fisher Information for Generalized Maxwell Model**

The elements of $J(\boldsymbol{\theta}_{MLE})_{i,j}$ for a generalized Maxwell model with $N$ relaxation times can be calculated following eqn. S10 as seen below.

$$J(G_{e_{MLE}}, G_{1_{MLE}}, \dots \tau_{N_{MLE}}) = \left[ - \begin{bmatrix} \partial^2_{G_e} & \partial_{G_e}\partial_{G_1} & \cdots & \partial_{G_e}\partial_{\tau_N} \\ \partial_{G_1}\partial_{G_e} & \partial^2_{G_1} & & \vdots \\ \vdots & & \ddots & \vdots \\ \partial_{\tau_N}\partial_{G_e} & \cdots & \cdots & \partial^2_{\tau_N} \end{bmatrix} L(\sigma_{obs}|G_e, G_1, \dots \tau_N) \right]_{G_{e_{MLE}}, G_{1_{MLE}}, \dots \tau_{N_{MLE}}} \quad (S11)$$

5. **Mechanical Models**

The constitutive equation for linear viscoelasticity is given as an arbitrary order differential equation which relates the stress and the strain of the material. Such an equation is generally solved by the Boltzmann integral which relates the stress $\sigma$ to a convolution of the viscoelastic modulus $Q$ and the strain $\varepsilon$[9–13]. In these types of equations, the modulus is referred to as the kernel of the integral and represents the impulse response of the material such that ( $Q(t) = Q(t) * \delta(t)$).

$$\sigma(t) = \int du \; Q(t-u)\varepsilon(u) \quad (S12)$$

Then, using the models derived by Lee and Radok, this equation can be equivalently obtained in terms of force $f$ and indentation $h$ between the viscoelastic half-space and a rigid indenter of various geometries as seen in eqn. S13[14]. The parameters $\alpha$ and $\beta$ depend on the geometry of the indenter and can be found in other sources[15–17].

$$\frac{f(t)}{\alpha} = \int du\ Q(t-u)h^\beta(u) \tag{S13}$$

### 6. Modified Fourier Transform

Recently, the Z-transform (ZT) was used to directly obtain $Q$ from force-indentation data[18,19]. The ZT can be thought of as a discrete analog to the Laplace transform. Recall that the Fourier transform of a function can be directly obtained from its Laplace transform by setting the complex Laplace domain variable $s = i\omega$. Here, the discrete Fourier transform (DFT) can similarly be obtained from the ZT, by setting $z = i\omega$[18,19]. In both of these cases, the real part of the complex variable is removed. If the real part were kept, the result would give the modified Fourier transform (MFT) (in the case of $s = a + i\omega$) and the modified discrete Fourier transform (MDFT) (in the case of $z = \ln(r_0) + i\omega$). As both the Z and Laplace transforms deal with the entirety of the complex plane, the modified transforms restrict this to a single line and circular contour, in the continuous and discrete cases, respectively. Thus, it is far more practical to use the MDFT to analyze the discrete force-indentation data.

$$T_{DFT}\{f[n]\}(\omega) = \sum_n f[n]e^{-i\omega n} \tag{S14}$$

$$T_{ZT}\{f[n]\}(z) = T_{ZT}\{f[n]\}(\omega, r) = \sum_n f[n]z^{-n} = \sum_n f[n]r^{-n}e^{-i\omega n} \tag{S15}$$

$$T_{MDFT}\{f[n]\}(\omega) = \sum_n f[n]r_0^{-n}e^{-i\omega n} = T_{DFT}\{f[n]r_0^{-n}\}(\omega) \tag{S16}$$

By applying the MDFT to eqn. S13, the convolution of $Q$ and $h^\beta$ turns into a multiplication of their transforms.

$$T_{MDFT}\left\{\frac{f[n]}{\alpha}\right\}(\omega) = T_{MDFT}\{Q[n]\}(\omega)\,T_{MDFT}\{h^\beta[n]\}(\omega) = Q(\omega)\,T_{MDFT}\{h^\beta[n]\}(\omega) \tag{S17}$$

This multiplication can then be directly manipulated to obtain $Q$ from the force-indentation data as seen in eqn. S18[19]. Of course, such a technique is not restricted to force-indentation data. The general definition of the Boltzmann integral can also be used, thus obtaining $Q$ from stress-strain data.

$$Q(\omega) = \frac{T_{MDFT}\left\{\frac{f[n]}{\alpha}\right\}(\omega)}{T_{MDFT}\{h^{\beta}[n]\}(\omega)} \tag{S18}$$

$$Q(\omega) = \frac{T_{MDFT}\{\sigma[n]\}(\omega)}{T_{MDFT}\{\varepsilon[n]\}(\omega)} \tag{S19}$$

While using the DFT may seem to be more convenient, doing so would result in an inaccurate characterization as the force-indentation data is numerically unbounded and does not represent a periodic signal[18,19]. The MDFT can be calculated from the DFT as seen in eqn. S16, though it should be noted that careful selection of $r_0$ is required to avoid errors in this inversion process[19]. It is typical for researchers to seek the storage and loss moduli of a material as the end goal of its viscoelastic characterization. These quantities are defined as the real and imaginary parts of the Fourier spectrum of $Q$. As the MDFT operates in the modified Fourier domain, the correspondence between the real and imaginary parts of $Q(\omega)$ obtained through this method do not exactly correspond to the real storage and loss moduli. Though they often are quite close, inaccuracies compound for excessively small and large values of $r_0$[18].

### 7. Dimensionless Model Derivation: Time Domain

The generalized Maxwell model is one of the simplest linear viscoelastic models that is capable of representing the canonical rheological behaviors of creep and stress relaxation. For these reasons, this model is a popular choice among those seeking to characterize viscoelastic material using force-indentation or stress-strain experiments.

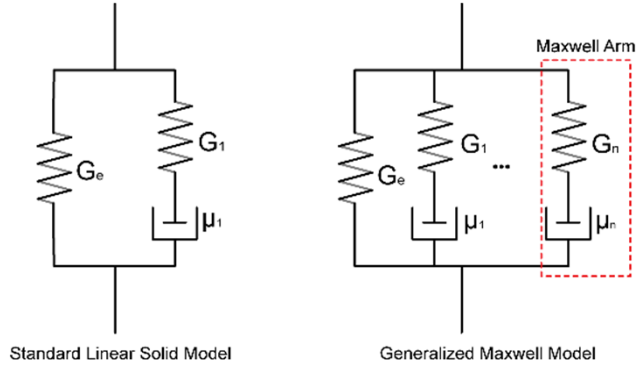

**Figure 1**: The Standard Linear Solid (SLS) model (seen on the left) is obtained as a special case of the generalized Maxwell model (seen on the right), where the number of relaxation times is set to 1.

Seen in Fig. 1, the generalized Maxwell model is comprised of a series of parallel 'Maxwell arms' made of an elastic spring with a modulus $G$ in series with a viscous dashpot with a viscosity $\mu$. The rate at which the stress in the $n^{th}$ Maxwell arm is governed by the ratio of $\mu_n/G_n$ and is referred to as the relaxation time $\tau_n$. The impulse response of such a material can be found from solving the stress-strain equation for the linear time invariant (LTI) system and is given as eqn. S20[9–11,13].

$$Q(t) = G_e \delta(t) + \sum_n^N \left[ G_n \delta(t) - \frac{G_n}{\tau_n} e^{-\frac{t}{\tau_n}} \right] \quad (S20)$$

The modified Fourier domain correspondence of $Q$ for this model can be obtained by solving the same equation using discrete signals and difference equations which results in eqn. 21[18,19]. It is an interesting note that in the limit of small $\omega$, eqn. S21 approaches the equivalent continuous Fourier domain correspondence.

$$Q(\omega) = G_e + \sum_{n}^{N}\left[G_n - \frac{G_n}{1 + \frac{\tau_n}{\Delta t}\left(1 - \frac{e^{-i\omega}}{r_0}\right)}\right] \tag{S21}$$

8. **Derivation of Dimensionless Models**

Due to the similarity in the structures of eqn.'s S12 and S13, we consider the quantities $f/\alpha$ and $h^\beta$ to be equivalent to stress and strain respectively. While the indentation $h$ in typical AFM force-indentation experiments typically does not perfectly follow a line, the 'strain', given as $h^\beta$, typically is closely linear as seen in Fig. 2.

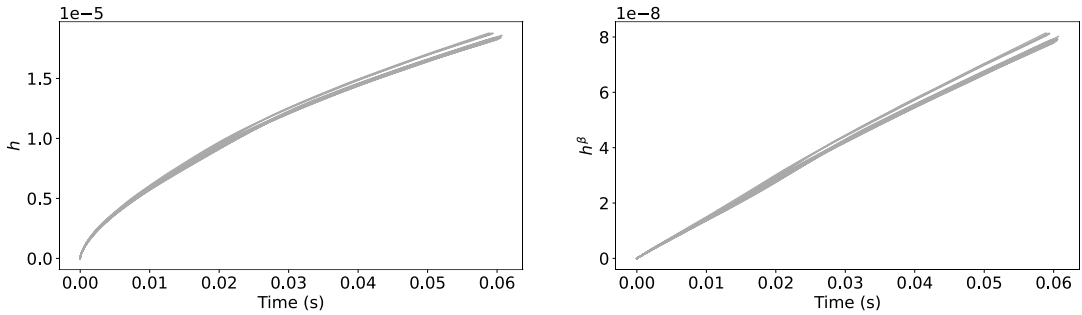

**Figure 2**: Indentation $h$ and 'strain' $h^\beta$ as functions of time obtained from an AFM experiment with a spherical probe ($\beta = 3/2$).

Then, assuming the 'strain' of a force-indentation experiment follows a linear function in time, we can write.

$$h^\beta(t) = \varepsilon_0\, t \tag{S22}$$

Substituting this linearized strain as well as the $Q(t)$ of a generalized Maxwell model, we can obtain the behavior of the stress / force for our idealized force-indentation curves.

$$\sigma(t) = \frac{f(t)}{\alpha} = \int du \left( G_e \delta(t-u) + \sum_n^N \left[ G_n \delta(t) - \frac{G_n}{\tau_n} e^{-\frac{t-u}{\tau_n}} \right] \right) \varepsilon_0 \, u \tag{S23}$$

Which can be solved exactly to yield eqn. S24.

$$\sigma(t) = \frac{f(t)}{\alpha} = \varepsilon_0 \left( G_e t + \sum_n^N G_n \tau_n \left( 1 - e^{-\frac{t}{\tau_n}} \right) \right) \tag{S24}$$

To isolate the behavior of the model parameters with respect to the experimental conditions, we can substitute the following normalizations where $T$ is the length of the experiment in time.

$$\tilde{G}_n = \frac{G_n}{G_e + \sum_n G_n} \tag{S25}$$

$$\tilde{\tau}_n = \frac{\tau_n}{T} \tag{S26}$$

$$\tilde{t} = \frac{t}{T} \tag{S27}$$

Performing these substitutions and further normalizing by $\varepsilon_0$ yields the following equation which is essential in obtaining the results in the main paper.

$$\tilde{\sigma}(\tilde{t}) = \frac{f(\tilde{t})}{\alpha \, \varepsilon_0 \, G_g \, T} = \tilde{t} + \sum_n^N \tilde{G}_n \tilde{\tau}_n \left( 1 - e^{-\frac{\tilde{t}}{\tilde{\tau}_n}} \right) - \tilde{G}_n \tilde{t} \tag{S28}$$

A similar approach can be taken to obtain the dimensionless form of $Q(\omega)$.

$$\tilde{Q}(\omega) = \frac{Q(\omega)}{G_g} = 1 - \sum_n^N \frac{\tilde{G}_n}{1 + \frac{\tilde{\tau}_n}{\Delta t}\left(1 - \frac{e^{-i\omega}}{1.001}\right)} \tag{S29}$$

Note that by returning to eqn. S24, it is evident that the timescales $t$ and $\tau_n$ are both scaled by the strain rate $\varepsilon_0$. As we mention in the main paper, this results in an equivalent nondimensional form of eqn. S28 where the dimensionless timescales are given by eqn.'s S26a and S27a. Interpreting these nondimensional parameters allows us to equivalently obtain an upper bound for obtainable relaxation times in terms of the strain rate. Specifically, to keep the dimensionless time constant $\tilde{\tau}_n$ in the information rich regime of the time domain experiments ($\tilde{\tau}_n < 0.1$), one requires that $\tau_n < 1/10\varepsilon_0$ thus allowing the specification of an optimal strain rate for the detection of specific relaxation times.

$$\tilde{\tau}_n = \tau_n \varepsilon_0 \tag{S26a}$$

$$\tilde{t} = t\varepsilon_0 \tag{S27a}$$

9. **Dimensionless Model Parameters**

As the arm moduli $\tilde{G}_n$ have been normalized by the instantaneous modulus ($G_g = G_e + \sum_n G_n$), they can obtain a minimum value of 0 which of course corresponds to a modulus of 0. The sum of all of the $\tilde{G}_n$'s can be, at most, 1 which corresponds to a material with $G_e = 0$, which is often interpreted as a fluid (i.e. the material will continue to relax its stress until it reaches a state of zero stress)[11]. The normalized relaxation times $\tilde{\tau}_n$ range between 0 and $\infty$, though we are mostly interested in their behavior within the observable timeframe of the experiment. Hence, we typically only consider $\tilde{\tau}_n$'s between 0 and 1. Similar to the relaxation times, the time axis $\tilde{t}$ has been normalized by the experiment length, thus restricting $\tilde{t}$ to be between 0 and 1 with 1 corresponding to the end of the experiment.

## 10. Surfaces of the $l^2$ Norm and Maximum Log Likelihood for the Dimensionless Models

The $l^2$ norm for the normalized time (eqn. S28) and frequency (eqn. S29) domain models can now be determined. Again, the $l^2$ norm is a function of the model parameters $\boldsymbol{\theta}$ and the successful identification of $\boldsymbol{\theta}$ from an optimization algorithm depends on its structure.

$$l_t^2(\boldsymbol{\theta}) = \sum_{\tilde{t}} [\tilde{\sigma}(\tilde{t}, \boldsymbol{\theta}) - \tilde{\sigma}_{obs}(\tilde{t})]^2 \tag{S30}$$

$$l_\omega^2(\boldsymbol{\theta}) = \sum_{\omega} \left[\tilde{Q}(\omega, \boldsymbol{\theta}) - \tilde{Q}_{obs}(\omega)\right]^2 \tag{S31}$$

We now perform parameter sweeps for these norms over a range of parameters for a model with a single relaxation time. In both cases, $\tilde{G}_1$ and $\tilde{\tau}_1$ are both varied between 0 and 1 and the resulting model predictions ($\tilde{\sigma}(\tilde{t}, \tilde{G}_1, \tilde{\tau}_1)$ and $\tilde{Q}(\omega, \tilde{G}_1, \tilde{\tau}_1)$) are compared to a simulated dataset with the parameters $\tilde{G}_{sim}, \tilde{\tau}_{sim}$. The resulting surfaces are plotted below.

In Fig. 3, we have $\tilde{G}_{sim} = \tilde{\tau}_{sim} = 0.5$ with these 'true' values marked with an X in the center of these surfaces. The left column of plots shows the surfaces of $l_t^2$ with an increasingly large amount of noise added to the simulated data. In the right column, the $l_\omega^2$ surfaces are shown, with the same increasing noise. As discussed in the main paper, the time domain contains a valley like formation with values below the machine precision level, thus indicating that optimization will likely fail in these areas. As the standard deviation of the noise increases, so too does the size of the valley. In comparison, the minimum in the frequency domain remains well defined even in the presence of considerable noise.

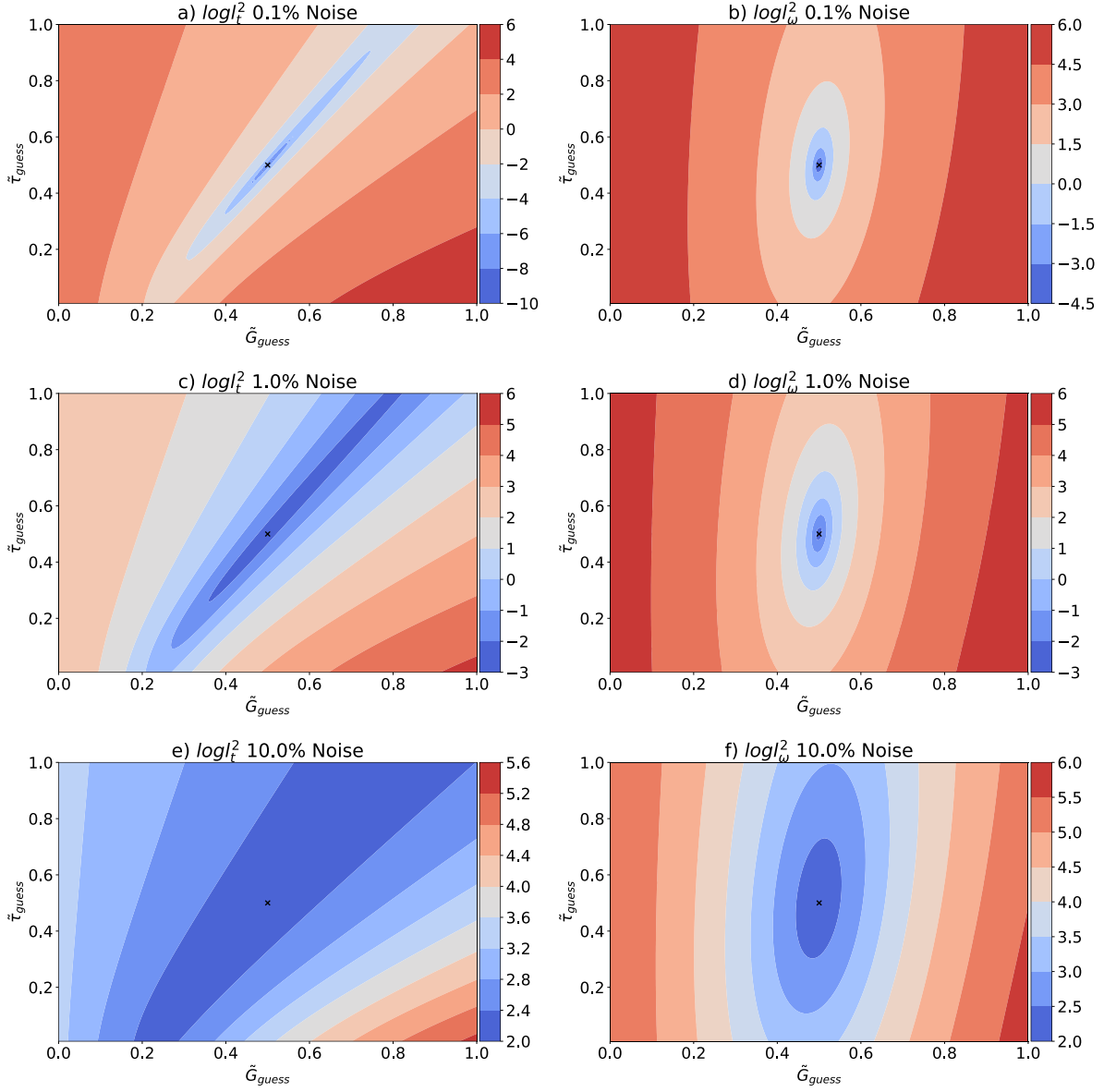

**Figure 3**: Surfaces of $l_t^2$ and $l_\omega^2$ for a range of parameter space. Compared to simulated dataset ($\tilde{G}_{sim} = \tilde{\tau}_{sim} = 0.5$) with an increasing amount of noise.

Next, we perform a similar calculation; however, the dataset is now simulated with a relaxation time $\tilde{\tau}_{sim} = 0.1$. As we have shown in the main paper, relaxation times that are less than $T/10$ are most readily obtained from the time domain. As expected, the size of the valley in $l_t^2$ is significantly reduced when compared to the previous case shown in Fig. 4. Furthermore, the expansion of the valley in $l_t^2$ is further restricted in this case – though it still indicates there will be considerable problems when optimizing.

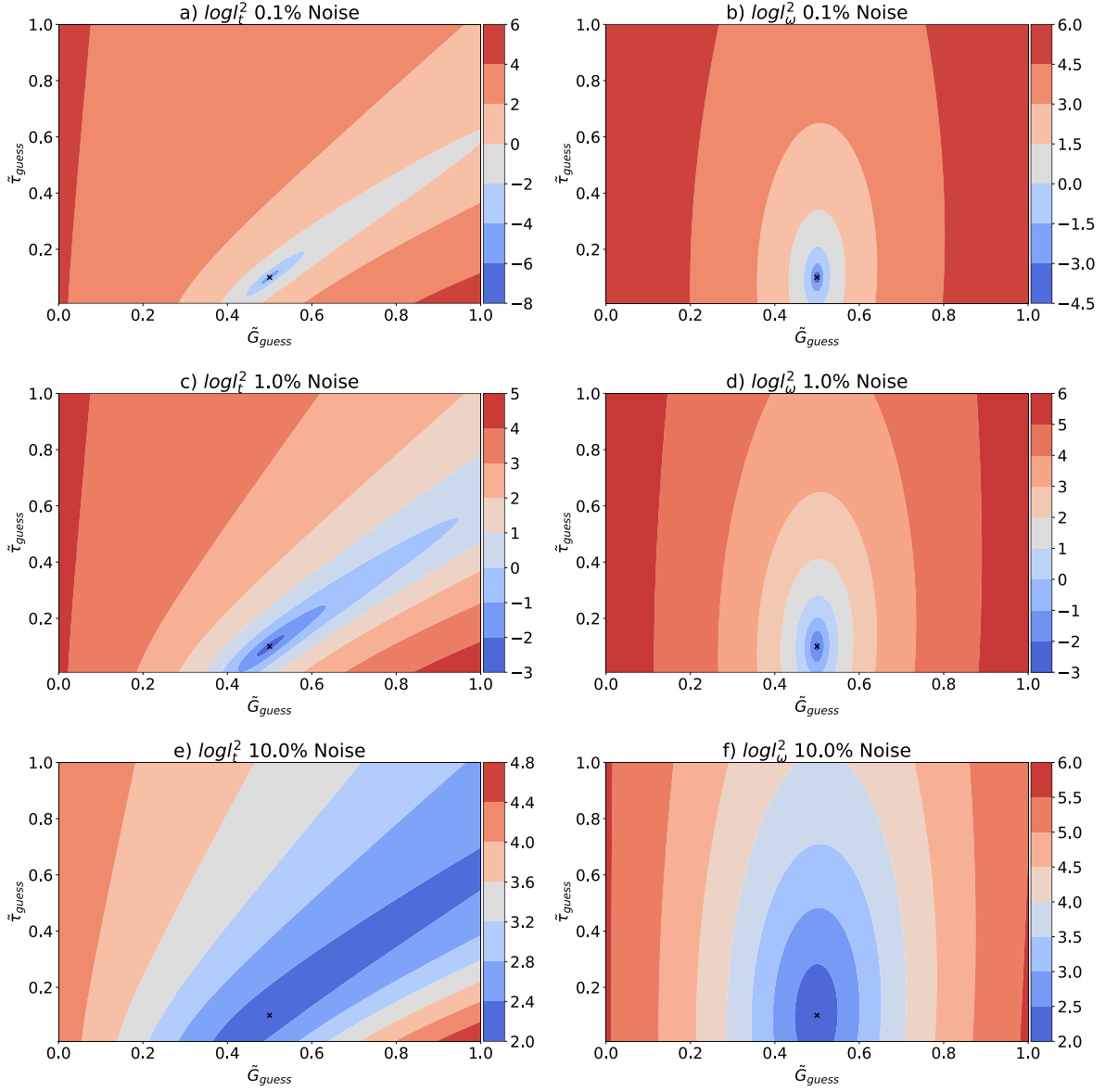

**Figure 4**: Surfaces of $l_t^2$ and $l_\omega^2$ for a range of parameter space. Compared to simulated dataset ($\tilde{G}_{sim} = 0.5$, $\tilde{\tau}_{sim} = 0.1$) with an increasing amount of noise.

As done in the main paper, we can further contrast these two approaches by using the log likelihood. Here, we seek to describe a material with two relaxation times by using a model with a single relaxation time (SLS model). In the two figures below, we provide the normalized log likelihood ($L(\tilde{\sigma}_{obs}|\boldsymbol{\theta}_{MLE})$ and $L(\tilde{Q}_{obs}|\boldsymbol{\theta}_{MLE}))$ for the maximum likelihood SLS estimate of various parameters ($\tilde{\tau}_1, \tilde{\tau}_2$ and $\tilde{G}_1, \tilde{G}_2$) of the two arm model in both the time (Fig. 5) and frequency (Fig. 6) domains.

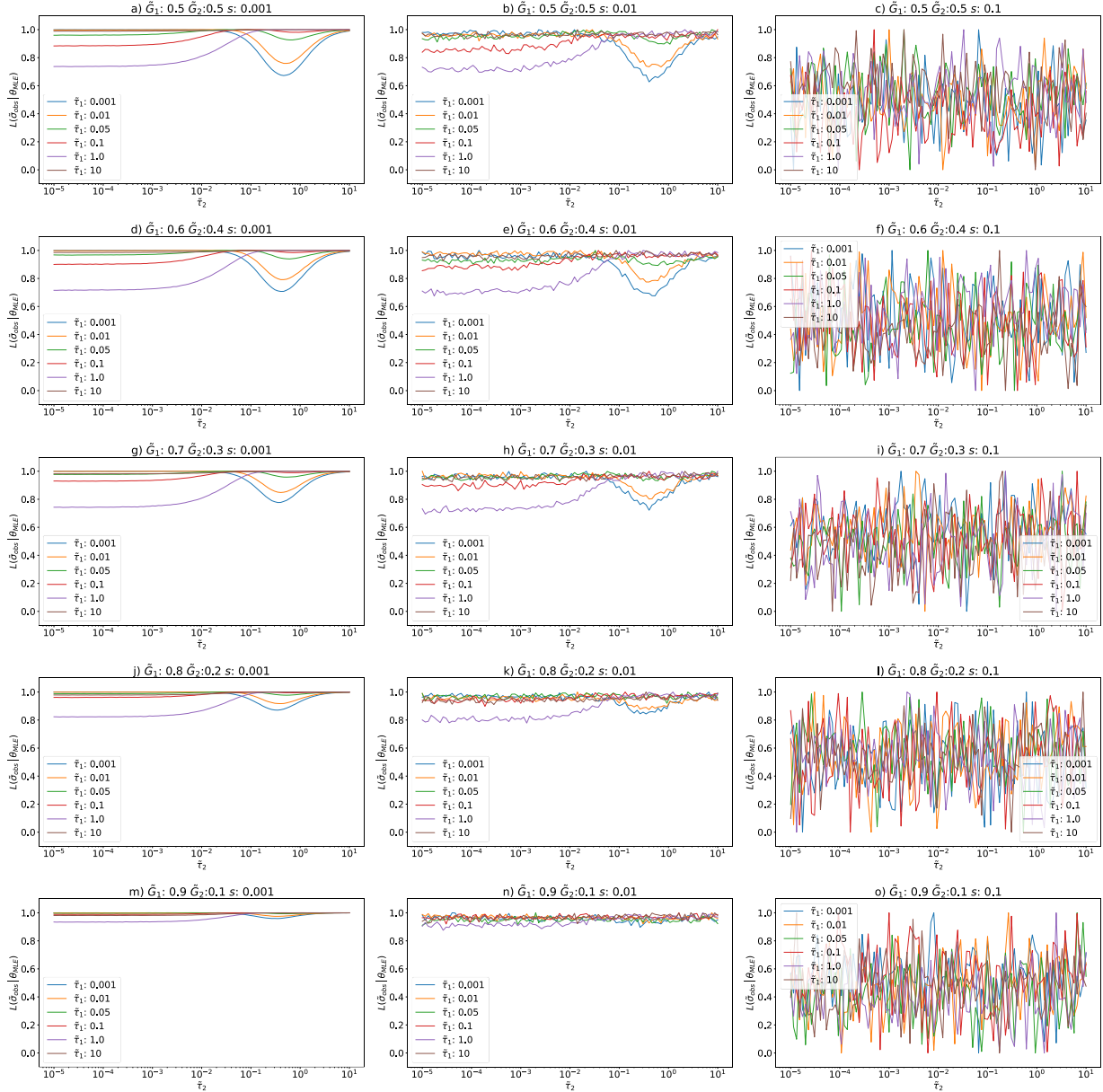

**Figure 5**: Values of the normalized maximum log likelihood estimate for an SLS model describing a material with two relaxation times $\tilde{\tau}_1, \tilde{\tau}_2$ and two moduli $\tilde{G}_1, \tilde{G}_2$ in the time domain. Moving from left to right, the columns correspond to data with noise of an increasing magnitude $s$. Each row corresponds to sets of values of $\tilde{\tau}_1, \tilde{\tau}_2$ and $\tilde{G}_1, \tilde{G}_2$ with the difference between $\tilde{G}_1$ and $\tilde{G}_2$ increasing from top to bottom.

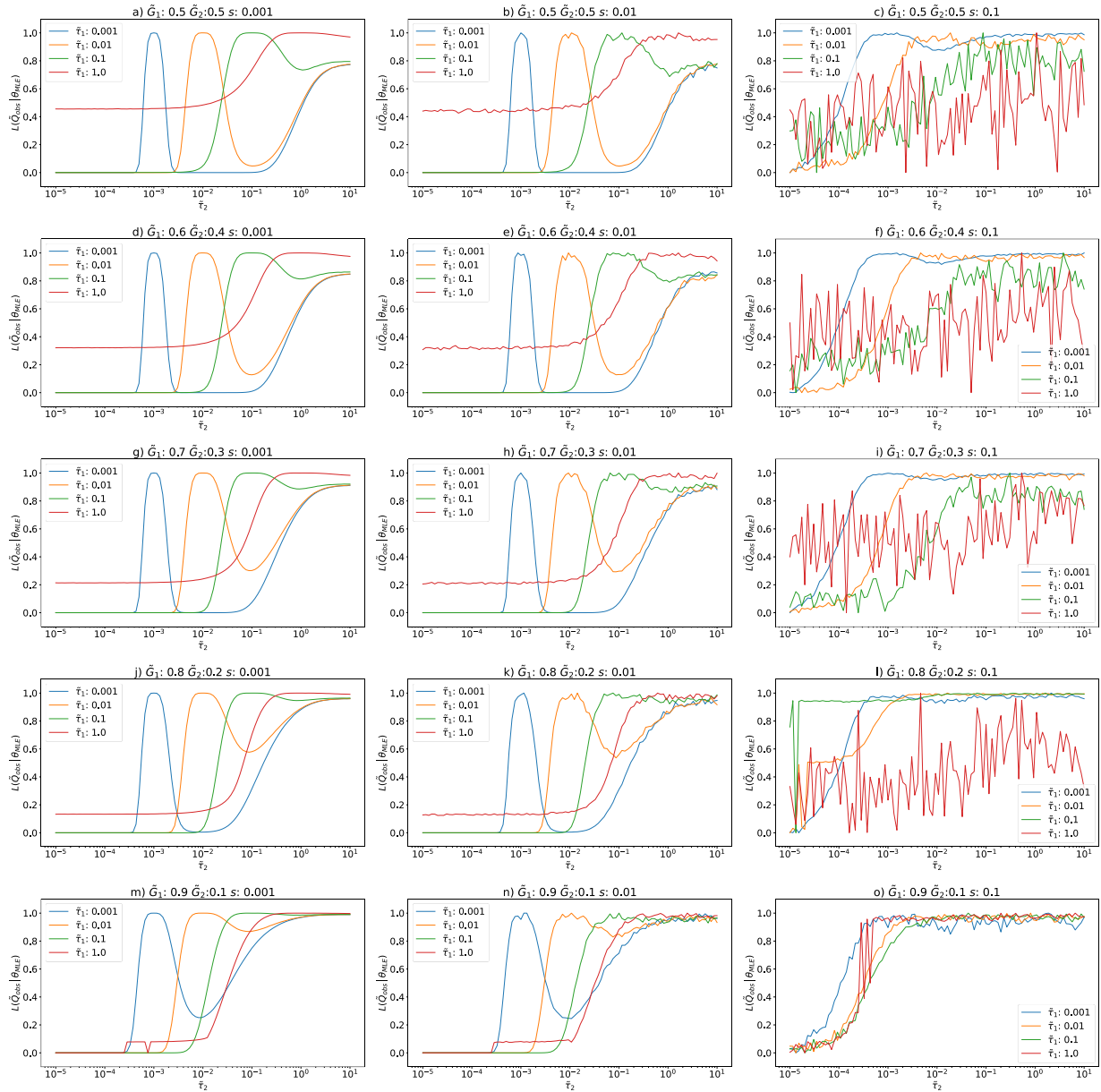

**Figure 6**: Values of the normalized maximum log likelihood estimate for an SLS model describing a material with two relaxation times $\tilde{\tau}_1, \tilde{\tau}_2$ and two moduli $\tilde{G}_1, \tilde{G}_2$ in the frequency domain. Moving from left to right, the columns correspond to data with noise of an increasing magnitude $s$. Each row corresponds to sets of values of $\tilde{\tau}_1, \tilde{\tau}_2$ and $\tilde{G}_1, \tilde{G}_2$ with the difference between $\tilde{G}_1$ and $\tilde{G}_2$ increasing from top to bottom.

In both cases, $L(\tilde{\sigma}_{obs}|\boldsymbol{\theta}_{MLE})$ and $L(\tilde{Q}_{obs}|\boldsymbol{\theta}_{MLE})$ form peaks where the values of $\tilde{\tau}_1$ and $\tilde{\tau}_2$ overlap. As discussed in the main paper, this is because as the two relaxation times come closer together, their behavior can be better approximated as a single relaxation time. Hence, for values of $\tilde{\tau}_1$ and $\tilde{\tau}_2$ that are close together, $L(\tilde{\sigma}_{obs}|\boldsymbol{\theta}_{MLE})$ and $L(\tilde{Q}_{obs}|\boldsymbol{\theta}_{MLE})$ will be maximized. With this in mind, we can determine that the superior method of characterization will be the one which offers the greatest distinction between relaxation times. We can see that as the width of the peaks in $L(\tilde{Q}_{obs}|\boldsymbol{\theta}_{MLE})$ are narrower than $L(\tilde{Q}_{obs}|\boldsymbol{\theta}_{MLE})$, the frequency domain offers a greater sensitivity for resolving multiple relaxation times from force-indentation experiments. A similar conclusion can be drawn from the decreased amplitude in $L(\tilde{\sigma}_{obs}|\boldsymbol{\theta}_{MLE})$ as compared to $L(\tilde{Q}_{obs}|\boldsymbol{\theta}_{MLE})$.

When the modulus associated with a relaxation time is decreased, the effect of that relaxation time is significantly reduced. In comparing Fig. 5a to 5m, for example, this effect is made obvious: $L(\tilde{\sigma}_{obs}|\boldsymbol{\theta}_{MLE})$ is effectively flattened for the case when $\tilde{G}_1$ is about an order of magnitude greater than $\tilde{G}_2$. Thus, determining $\tilde{\tau}_2 = 0.1s$ when $\tilde{\tau}_1 = 0.001s$ for $\tilde{G}_1 = 0.9$ and $\tilde{G}_2 = 0.1$ is as difficult as determining $\tilde{\tau}_2 = 0.01s$ when $\tilde{\tau}_1 = 0.001s$ for $\tilde{G}_1 = \tilde{G}_2 = 0.5$. However, these issues are significantly reduced in the frequency domain as seen in Fig. 6a and 6m, for example.

### 11. Discrete Convolution Error

It is uncommon for researchers to assume that the indentation in an AFM experiment follows a prescribed functional form as we have done in the derivation of the linearized model. In practice, the convolution definition given in eqn. S13 is used more frequently as it accommodates force and indentation data of almost any arbitrary form, without the need for assuming one as we have done here. However, as the experimental data is inherently discretized, one must use a *discrete* convolution operation rather than a *continuous* one. In this case, such an approximation is only valid when the convolved functions are band limited (i.e. their frequency components lie within a finite region of the Fourier domain)[20]. Of course, one may be able to perform experiments which use band limited strain/indentation signals such as sines or cosines; however, most signal shapes used in force-indentation experiments are not exactly so. Though, by admission, the family of ramp-like signals that are more commonly used in these experiments approximately satisfy this criterion due to their rapid attenuation in the frequency domain. However, this issue still remains for the viscoelastic modulus. In particular, band limited moduli may not correspond to physical behaviors of materials.

To address this question as generally as possible, we should first consider the general shape of the viscoelastic modulus in the frequency domain. The behavior of such functions near the origin either starts from zero, corresponding to a fluid, or a finite nonzero value, corresponding to a solid. Following this, the function tends towards some nonzero value at high frequency. This general behavior can be represented as a superposition of a positive constant (DC shift), representing the immediate response of the material, with some function of frequency, representing the transient response of the material. Indeed, such decompositions are commonplace in the general theory of linear viscoelasticity[12,13]. In this case, the frequency components associated with the positive constant result in a Dirac delta function in the time domain, which is analytically equivalent to a scaling of the strain\indentation input with the positive constant. The remaining transient portion of the modulus must then be convolved with the input which is where the issue of band limiting arises.

To further assess these issues, we must assume that the transient behavior follows that of a generalized Maxwell model. For this case, the constitutive equation is again decomposed into immediate and transient components as discussed above which results in eqn. S33. A brief inspection of the spectrum of $e^{-t/\tau_n}$ in Fig. 5 reveals that it has a decaying behavior as the magnitude of the frequency increases. Thus, we may posit that approximating $e^{-t/\tau_n}$ as being band limited may not introduce significant errors in the calculation of the convolution. To fully determine if this is true, we first can calculate the spectrum of $e^{-t/\tau_n}$ given as $q$ as seen below.

$$\frac{f(t)}{\alpha} = \int du \left( G_e \delta(t-u) + \sum_n^N \left[ G_n \delta(t) - \frac{G_n}{\tau_n} e^{-\frac{t-u}{\tau_n}} \right] \right) h^\beta(u) \tag{S32}$$

$$\frac{f(t)}{\alpha} = \left( G_e + \sum_n^N G_n \right) h^\beta(t) - \sum_n^N \frac{G_n}{\tau_n} \int du \ e^{-\frac{t-u}{\tau_n}} h^\beta(u) \tag{S33}$$

$$q(\omega) = \frac{1}{1 + i\omega\tau} \tag{S34}$$

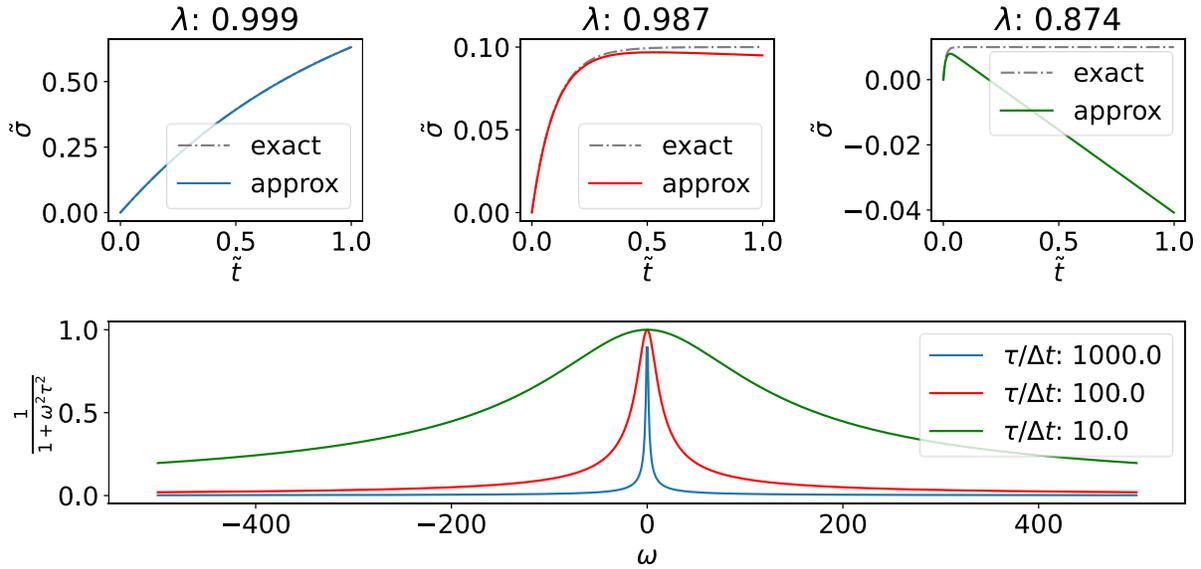

Figure 7: Three force-indentation curves calculated exactly (shown in grey) and approximately using the discrete convolution (shown in color). The percent of the total energy of the spectrum contained in the Nyquist band limits is included to demonstrate how much energy is needed to be captured in order for this approximation to be valid. The magnitudes of the spectra for these three force-indentation curves

are also included, demonstrating that larger relaxation times relative to the sampling timestep cause the spectra to be better approximated as band limited.

As band limited spectra have their entire energy contained within a finite frequency range, we can assess the ability of $q$ to be approximated as a band limited spectrum by looking at the amount of its energy that is contained in the Nyquist frequency bands ($-1/2\Delta t$ to $1/2\Delta t$). For $q$ to be determined to be approximately band limited, it should then have most of its energy contained in this range. To begin, we first determine the energy of $q$.

$$|q(\omega)| = \frac{1}{1 + \omega^2 \tau^2} \tag{S35}$$

Then, assuming that the spacing between the frequencies are sufficiently small (i.e., the experiment is sufficiently long), the total energy contained between the frequencies $-\omega_n$ and $\omega_n$ can be obtained as an integral of $|q(\omega)|$.

$$E(\omega_n) = \int_{-\omega_n}^{\omega_n} d\omega \frac{1}{1 + \omega^2 \tau^2} \tag{S36}$$

As previously mentioned, a band limited signal should have 100% of its energy contained in a given frequency range, say $-\omega_n$ and $\omega_n$. Thus, the ratio $\lambda$ of the energy in this frequency range to the energy in all the frequencies must be 1, or as close to it as possible for the approximate case.

$$\lambda = \frac{E(\omega_n)}{E(\infty)} = \frac{E\left(\frac{1}{2\Delta t}\right)}{E(\infty)} \tag{S37}$$

These energies can then be calculated from the integral in eqn. S36.

$$\lambda = \frac{2\,\text{atan}\left(\frac{\tau}{2\Delta t}\right)}{\pi} \tag{S38}$$

Finally, to maximize $\lambda$, we obtain an inequality between $\tau$ and $\Delta t$ which gives the final eqn. S39.

$$\tau > 2\,\Delta t\,\tan\left(\frac{\lambda\pi}{2}\right) \tag{S39}$$

As demonstrated in Fig. 5, the convolution works best for values of $\lambda > 0.999$ which corresponds to roughly $\tau > 1000\,\Delta t$. Therefore, when using the convolution definition of the generalized Maxwell model, one can only use relaxation times that are thousands of times larger than the sampling timestep. Thus, an even more restrictive lower limit can be placed on the relaxation times that can be determined in the time domain if the convolution definition is used.

## 12. Noise Carried Through Spectral Inversion with MDFT

As mentioned in Section 6, the viscoelastic modulus of a material can be directly obtained from force-indentation indentation by applying the modified discrete Fourier transform (MDFT) to the data. Once transformed, the force and indentation spectra can be divided to obtain the modulus; however, this division can magnify the effects of even the slightest noise in the data. For example, Fig. 6 shows the spectra of the force and indentation from an AFM experiment. Although the individual spectra appear to be free from noise, the resulting viscoelastic modulus is severely distorted at high frequencies.

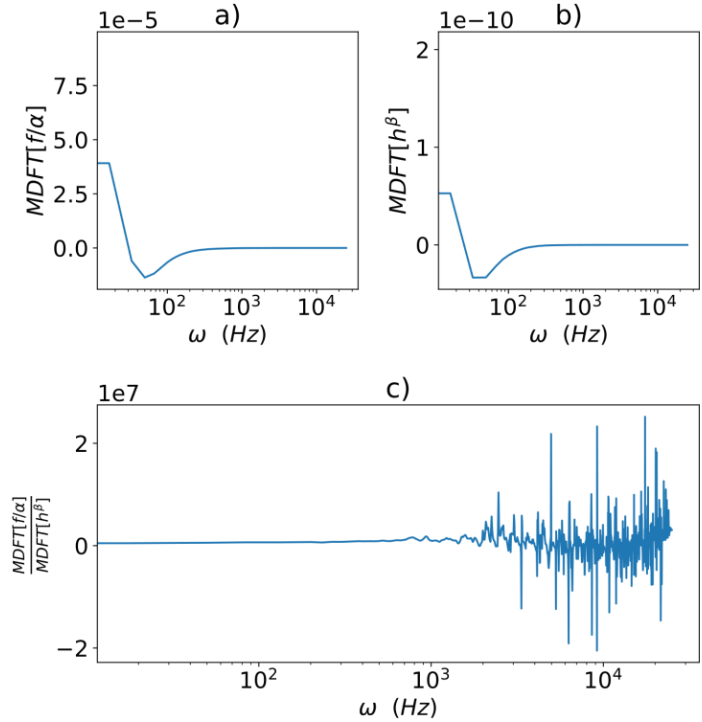

**Figure 8**: Spectra of a) force and b) indentation data from an AFM experiment obtained from the MDFT as described in Section 6. Though the individual spectra seem noticeably free from noise, dividing them to obtain the c) viscoelastic modulus magnifies the high frequency noise.

To understand how this noise manifests in the frequency domain, we use the more generalized stress-strain relationship from eqn. S12. Here, consider measuring the stress and strain of a material in an experiment where the data is polluted with noise $x$ which is generated by a mean zero Gaussian process with a standard deviation $s$. The resulting stress-strain equation will be given as eqn. S40.

$$\sigma(t) + x(t) = \int du \ Q(t-u) \left(\varepsilon(u) + x(u)\right)$$
(S40)

As the data has been polluted with noise, the value of $Q$ obtained from inverting eqn. S40 will be incorrect, hence we denote it as $Q'$. The result can be seen in eqn. S41 where the notation $f_\omega$ is used to denote the MDFT of some signal $f$.

$$\sigma_\omega + x_\omega = Q_\omega(\varepsilon_\omega + x_\omega) \tag{S41}$$

$$Q'_\omega = \frac{\sigma_\omega + x_\omega}{\varepsilon_\omega + x_\omega} \tag{S42}$$

However, the resulting equation is invalid as the spectrum of a random process such as $x$ is undefined. Instead, we will use the power spectrum of $Q'_\omega$[21].

$$|Q'_\omega|^2 = \frac{\sigma_\omega + x_\omega}{\varepsilon_\omega + x_\omega}\left(\frac{\sigma_\omega^* + x_\omega^*}{\varepsilon_\omega^* + x_\omega^*}\right) = \frac{|\sigma_\omega|^2 + |x_\omega|^2 + \sigma_\omega^* x_\omega + \sigma_\omega x_\omega^*}{|\varepsilon_\omega|^2 + |x_\omega|^2 + \varepsilon_\omega^* x_\omega + \varepsilon_\omega x_\omega^*} \tag{S43}$$

As the power spectrum of a random variable is its autocorrelation in the time domain, we can then substitute the variance $s^2$ of $x$ for $|x_\omega|^2$.

$$|Q'_\omega|^2 = \frac{|\sigma_\omega|^2 + s^2 + \sigma_\omega^* x_\omega + \sigma_\omega x_\omega^*}{|\varepsilon_\omega|^2 + s^2 + \varepsilon_\omega^* x_\omega + \varepsilon_\omega x_\omega^*} \tag{S44}$$

To completely remove the dependence of $Q'_\omega$ on the random variables, we take the expected value. To do so, we first need the following identity of the expected value of the spectra of $x_\omega$ and $x_\omega^*$.

$$E(x_\omega) = E\left[\sum_n x[n] r^{-n} e^{-i\omega n}\right] = \sum_n r^{-n} e^{-i\omega n} E[x[n]] = 0 \tag{S45}$$

$$E(x_\omega^*) = E\left[\sum_n x[n] r^{-n} e^{i\omega n}\right] = \sum_n r^{-n} e^{i\omega n} E[x[n]] = 0 \tag{S46}$$

Thus, the expected value of the magnitude of the modulus is given below.

$$E(|Q'_\omega|^2) = \frac{|\sigma_\omega|^2 + s^2}{|\varepsilon_\omega|^2 + s^2} \tag{S47}$$

To understand how this erroneous modulus behaves with respect to the amplitude of the noise in the data, we can take a Maclaurin expansion of eqn. S47 to obtain a simple scaling behavior which converges for all values of $s^2 < |\varepsilon_\omega|^2$ (i.e. reasonable amounts of noise).

$$E(|Q'_\omega|^2) = \frac{|\sigma_\omega|^2}{|\varepsilon_\omega|^2} + \sum_{n=1}^{\infty} (-1)^n \frac{|\sigma_\omega|^2 - |\varepsilon_\omega|^2}{(|\varepsilon_\omega|^2)^{n+1}} s^{2n} \tag{S48}$$

Since the division of $|\sigma_\omega|^2$ and $|\varepsilon_\omega|^2$ gives the true value of $|Q_\omega|^2$, we can rewrite eqn. S48 to give the following, which is interpreted as the expected value of the error in obtaining $Q_\omega$ due to noise.

$$E(|Q'_\omega|^2 - |Q_\omega|^2) = \delta_Q = \sum_{n=1}^{\infty} (-1)^n \frac{|Q_\omega|^2 - 1}{|\varepsilon_\omega|^{2n}} s^{2n} \tag{S49}$$

Thus, we deduce that the error in the modulus scales according to the rule in eqn. S50.

$$\delta_Q \sim \frac{s^2}{|\varepsilon_\omega|^2} \tag{S49}$$

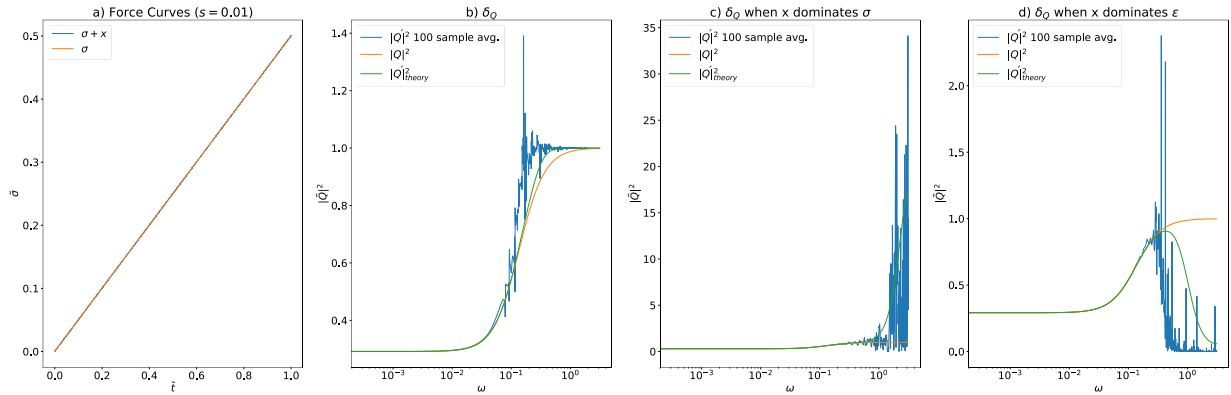

**Figure 9**: a) force curves with and without noise shown in the blue and orange curves, respectively. Even in the case where the amplitude of the noise is small relative to the full scale of the data, the b) error seen when inverting the force curves to obtain the viscoelastic modulus is large. Here, a 100-sample average (shown in blue) was taken to obtain an approximate expected value for this behavior. Two more plots are included which demonstrate the effect of noise when it dominates the c) stress signal and d) the strain signal.

As seen previously in Fig. 6, the spectra of typical AFM force and indentation data strongly decay at high frequencies. Thus, one can expect that since the error has an inverse dependence on the magnitude of the strain, the error will increase with frequency as seen in Fig. 7b. Most noticeable is the fact that even though the amplitude of the noise used in this example is small, the expected inverted spectrum (green) noticeably deviates from the true behavior (orange). We then perform a 100-sample average to gain a practical determination of this expected value as seen in the blue curve. While the effects of the noise in this average are even more pronounced than expected, the spectral averaging considerably reduces the error when compared to individual, non-averaged spectra. Further examples have been provided in Fig. 7c and 7d where the noise is assumed to dominate the stress only (Fig. 7c) and strain only (Fig. 7d). In these cases, the error due to the noise is considerably worse.

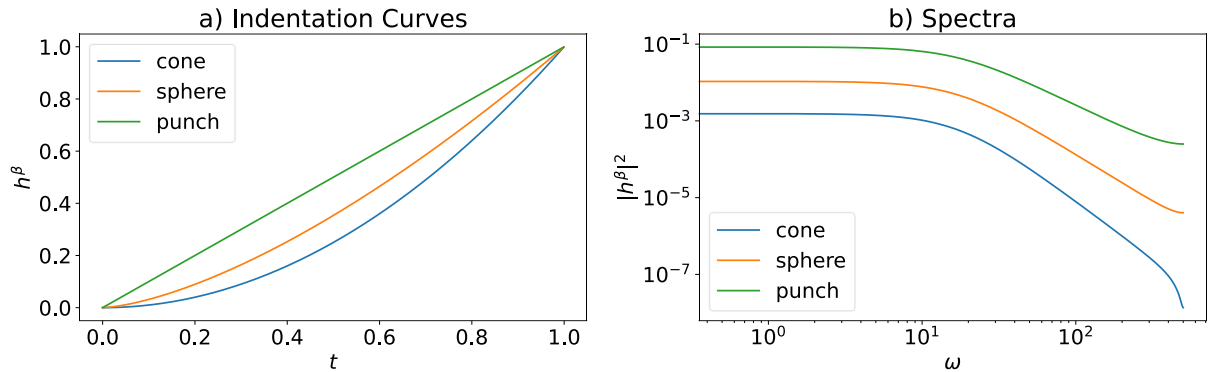

**Figure 10**: a) 'strain' curves $h^\beta$ generated for linear indentations for different indenter profiles (flat punch $\beta = 1$, sphere $\beta = 3/2$, and cone $\beta = 2$) and b) the magnitude of their spectra in the modified Fourier domain.

Fig. 10 demonstrates various 'strain' profiles given by $h^\beta$ for a linear indentation. Here, the different profiles for different indenter geometries are included as well as the magnitudes of their spectra. We see that the conical profile has a spectrum that decays most rapidly whereas the punch decays the slowest. Thus, for the same indentation form, a conical indenter will have the most error in the presence of noise whereas the flat punch will have the least. More generally, less divergent excitations will have a less rapid attenuation of their magnitude in the frequency domain. Therefore, we can expect that experiments with more gradual excitations will have less high frequency noise than those with more rapid, aggressive ones.